\RequirePackage{fix-cm}
\documentclass[epjc3,english,pdftex,twocolumn]{svjour3}
\journalname{Eur. Phys. J. C}
\renewcommand{\makeheadbox}{} 

\usepackage[english]{babel}
\usepackage[colorlinks=true,citecolor=blue,filecolor=blue,linkcolor=blue,urlcolor=blue,pdftex]{hyperref}
\usepackage[numbers,sort&compress]{natbib}

\usepackage{graphicx}%
\usepackage{multirow}%
\usepackage{amsmath,amssymb,amsfonts}%
\usepackage{mathrsfs}%
\usepackage[title]{appendix}%
\usepackage{xcolor}%
\usepackage{textcomp}%
\usepackage{manyfoot}%
\usepackage{booktabs}%
\usepackage{dcolumn}
\usepackage{algorithm}%
\usepackage{algorithmicx}%
\usepackage{algpseudocode}%
\usepackage{listings}%
\usepackage{siunitx}
\usepackage{comment}
\usepackage{setspace}

\begin{document}

\title{Characterisation of the Bedretto Underground Site for Fundamental Physics Experiments}

\author{Bj\"orn Penning\thanksref{e1,addr1}\and
        Nicolas Angelides\thanksref{addr1}\and
        Laura Baudis\thanksref{addr1}\and
        Harvey Birch\thanksref{addr1}\and
        Abigail Flowers\thanksref{addr1}\and
        Florian J\"org\thanksref{e2,addr1}\and
        Alexander Kavner\thanksref{e3,addr1}\and
        Marcelle Soares-Santos\thanksref{addr1}\and
        Aravind Remesan Sreekala\thanksref{addr1}\and
        Johannes W\"uthrich\thanksref{addr1}\and
        Guandi Zhao\thanksref{addr1,addr3}\and
        Chiara Capelli\thanksref{addr1}\and
        John Clinton\thanksref{addr6}\and
        Jose Cuenca Garc\'ia \thanksref{addr1}\and
        Paolo Crivelli\thanksref{addr3}\and
        Domenico Giardini\thanksref{addr2}\and
        Evangelos-Leonidas Gkougkousis\thanksref{addr1}\and
        Yacine Haddad\thanksref{addr5}\and
        Marian Hertrich\thanksref{addr2}\and
        Rebecca Hochreutener\thanksref{addr2}\and
        Luisa H\"otzsch\thanksref{addr1}\and
        Philippe Jetzer\thanksref{addr1}\and
        Ben Kilminster\thanksref{addr1}\and
        Boris Korzh\thanksref{addr4}\and
        Frédérick Massin\thanksref{addr6} \and 
        Knut Dundas Mor\aa\thanksref{addr1}\and
        Margherita Noia\thanksref{addr1}\and
        Francesco Piastra\thanksref{addr1}\and
        Christian Regenfus\thanksref{addr3}\and
        Federico Sanchez\thanksref{addr4}\and
        Steven Schramm\thanksref{addr4}\and
        Francesco Riva\thanksref{addr4}\and
        Serhan Tufanli\thanksref{addr5}\and
        Michele Weber\thanksref{addr5}\and
        Stefan Wiemer\thanksref{addr2}\and
        Mathilde Wimez\thanksref{addr2}
}
\thankstext{e1}{e-mail: bjoern.penning@uzh.ch}
\thankstext{e2}{e-mail: florian.joerg@physik.uzh.ch}
\thankstext{e3}{e-mail: alexander.kavner@physik.uzh.ch}

\institute{Physik-Institut, University of Z\"urich, Winterthurerstrasse  190, Z\"urich, 8057, Switzerland \label{addr1}
           \and
           Bedretto Underground Laboratory, ETH Z\"urich, Sonneggstrasse 5, Z\"urich, 5092, Switzerland \label{addr2}
           \and
           Departement Physik, ETH Z\"urich, Otto-Stern-Weg 1, Z\"urich, 8093, Switzerland \label{addr3}
           \and
           Departement de Physique Nucléaire et Corpusculaire, University of Geneva, 24, quai Ernest-Ansermet, Geneva, 1205, Switzerland \label{addr4}
           \and
           Physikalisches Institut, University of Bern, Sidlerstr. 5, Bern, 3012, Switzerland \label{addr5}
           \and
           Swiss Seismological Service (SED), ETH Zürich, Sonneggstrasse 5, Zürich, 8092, Switzerland \label{addr6}
}

\maketitle

\abstract{
Underground laboratories provide the ultra-low background and low-vibration environments essential for rare-event searches, gravitational-wave detection, and quantum-sensing technologies. We report a comprehensive environmental characterisation of the Be\-dretto tunnel in Ticino, Switzerland, a site offering horizontal access, excellent infrastructure, and the potential to be Europe’s second-deepest and quietest underground laboratory. At the prospective physics site, located beneath an overburden exceeding 1400\,m, we measure the cosmic-muon, $\gamma$-ray, and neutron fluxes, as well as the radon concentration, magnetic-field spectrum, and seismic backgrounds. The muon flux is suppressed by six orders of magnitude relative to the surface, consistent with an effective depth of about 4000\, metre water equivalent, gamma-ray and neutron measurements reflect the local geology and guide shielding requirements for future particle and nuclear physics experiments. Magnetic and seismic noise levels are found to be exceptionally low, meeting or exceeding the criteria for next-generation atom-interferometric gravitational-wave detectors. These results establish the site as a highly competitive, accessible deep-underground location for fundamental-physics experiments.
}




\section{Introduction}\label{sec1}
Underground laboratories provide uniquely background-free environments and are indispensable tools to understand fundamental physics processes. The massive rock overburden shields effectively even from the highest energetic cosmic rays, they are seismically very quiet since most vibrations travel at the surface and are attenuated quickly at depth, and  
Newtonian noise is reduced by the very large mass of rock surrounding the laboratory. Examples of underground experiments include rare-event searches such as the search for dark matter and neutrinoless double beta decay, rare-decay nuclear physics measurements, low-background material assay, environmental monitoring, and optimal sensitivity and stability of quantum sensors. Furthermore, the stable, low-background, and low-vibration environment offered by underground locations is highly beneficial for gravitational wave detection. Several major deep underground laboratories exist, but none in Switzerland. These facilities often have to prioritise large-scale experiments and offer only limited space for R\&D setups and prototyping. Furthermore, their often remote locations and necessary travel lead to logistical and practical overhead. The Bedretto site in the Canton of Ticino, Switzerland, provides an ideal location to establish a Swiss deep underground laboratory for fundamental physics. As one of the deepest and quietest laboratories in Europe, it is easily accessible from all Swiss research institutions and features horizontal access to the underground site.

The Bedretto tunnel is an annex to the Furka railway starting in Bedretto, Ticino, Switzerland. The tunnel was used to remove excavation material during construction and as an evacuation route for the railway. 

The tunnel presently hosts the Bedretto Underground Laboratory for Geosciences and Geoenergies (BedrettoLab), operated by ETH Z\"urich. The laboratory infrastructure, including electrical power, network connectivity, ventilation, and safety equipment, is well established. 
    
\begin{figure}[!h]
        \centering
        \includegraphics[width=0.99\linewidth]{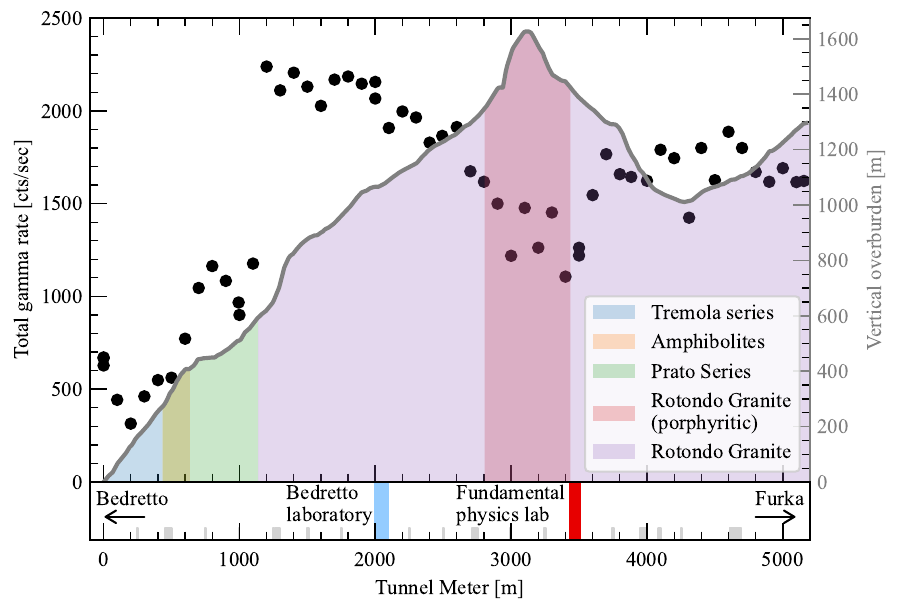}
        \caption{The Bedretto tunnel and the profile of the mountain with the different rock formations overlaid\,\cite{rast2022geology,nasa2019aster}. The black markers indicate the gamma-ray rate as measured using a 3~in NaI(Tl) detector. The red line indicates the site of the planned fundamental physics lab. Serendipitously, the laboratory's intended location, the gamma-ray flux is close to minimal while the overburden is maximal. The locations of rock niches are indicated in grey in the bottom panel.}
        \label{fig:tunnel_gamma}
    \end{figure}

The total length of the tunnel is $5220$\,m with a width and height of $2.70$\,m and a maximum overburden of about 1632\,m~\cite{Bedretto,bedretto_charaterization}. The intended site for the underground physics laboratory, as well as a primary focus of the measurements presented in this work, is located about 3.5\,km into the tunnel from the surface entrance and is denoted TM3500. The site is nearly underneath the peak of Pizzo Rotondo with an estimated direct vertical overburden of more than $1400$\,m~\cite{swisstopo}. The Bedretto tunnel crosses through three geological units of the Gotthard massif, the Tremola series, the Prato series, and the Rotondo granite~\cite{rast2022geology}, as shown in Fig.~\ref{fig:tunnel_gamma}. 

    To determine the suitability of the site for low-background physics, a comprehensive site characterisation campaign was performed in summer and autumn 2025. We conducted measurements of key background contributions such as radiative backgrounds, specifically cosmic muon, neutron, and $\gamma$-ray fluxes, and radon activity, as well as magnetic and seismic background levels.

\section{Atmospheric Muon Flux}

\begin{figure}[!h]
    \centering
        \includegraphics[width=1\linewidth]{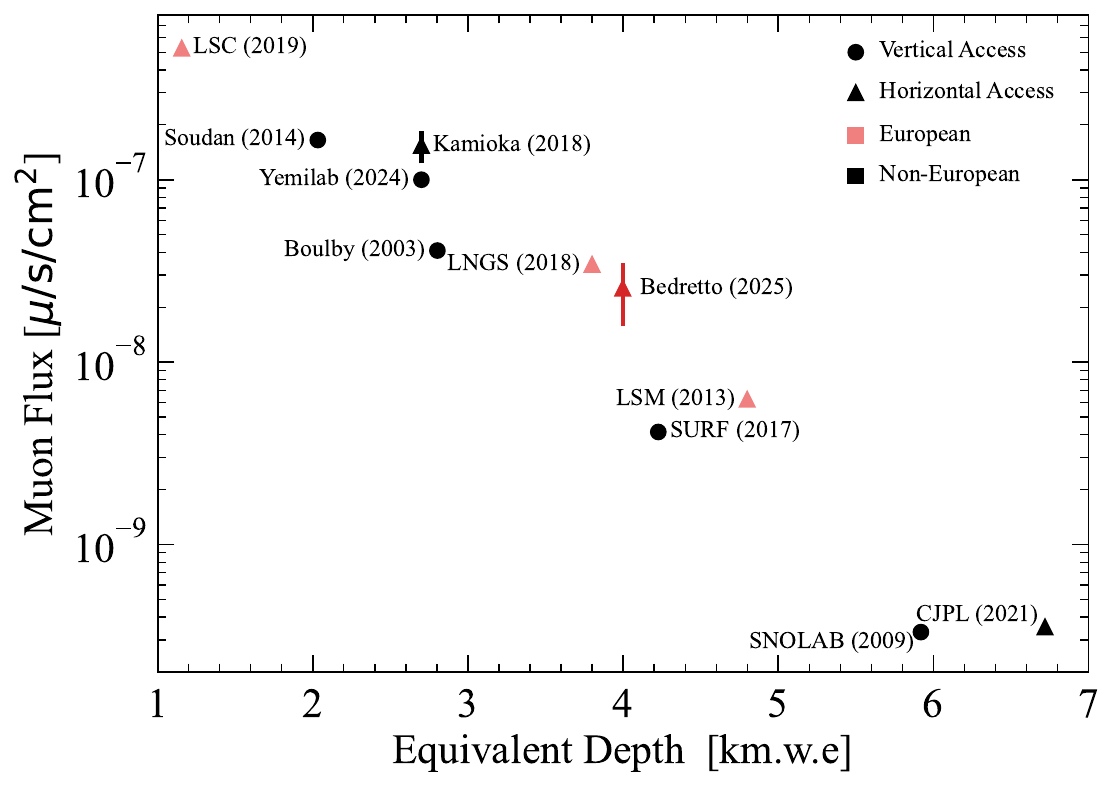}
        \caption{Measured muon flux rates and the equivalent water depth for different underground laboratories. The muon flux in Bedretto is suppressed by about a factor of $10^{6}$,  better than almost all existing underground laboratories in Europe designated in red. Muon fluxes taken from Refs.~\cite{Woodley:2024eln,Soudan_Muon,Yemilab_Muon,Boulby_Muon,LSC_Muon,LNGS_Muon,LSM_Muon,SURF_Muon,CJPL_Muon,SNO_Muon}.}
        \label{fig:waterdepth}
    \end{figure}

    Muons and neutrons from muon-induced spallation processes constitute a major and difficult to shield background for rare event searches.  As minimum-ionising particles, high-energy atmospheric muons are able to penetrate large amounts of matter, and only underground laboratories allow for a significant reduction of the cosmic muon flux. Underground laboratories are typically characterised in terms of metre-water-equivalent (m.w.e) in their ability to attenuate the cosmic muon flux~\cite{Heise:2022iaf,Bruno:2019vad,Woodley:2024eln,LSM_Muon}, see Fig.~\ref{fig:waterdepth}.

    \subsection{Muon Telescope}
    The atmospheric muon flux was measured at the TM3500 site utilising a telescope composed of four EJ-200 plastic scintillator panels~\cite{EJ-200}, each with an area of 625\,cm$^{2}$ and 3.8\,cm in thickness. An array of four \textit{Onsemi C-series} silicon photomultipliers (SiPM) was attached to each panel, biased with a \textit{CAEN} SiPM power supply at 31\,V~\cite{CAENsipm} and read out with a \textit{CAEN DT5730} 500\,MHz digitizer employing DPP-PHA firmware~\cite{CAEN5730,CAEN-DPP-PHA}. Lead plates, about $2$\,cm thick, were placed between the plastic scintillator to minimise spurious gamma and Compton scattering backgrounds. 

\begin{figure}[ht]
        \centering
        \includegraphics[width=1\linewidth]{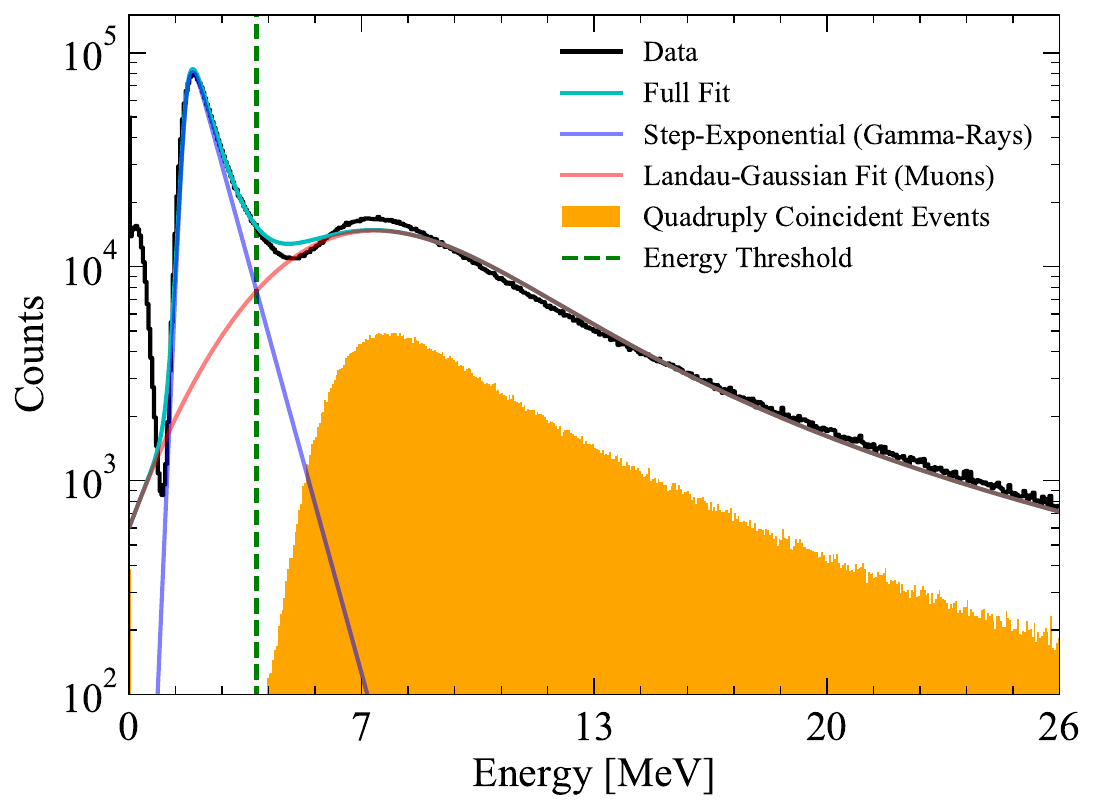}
        \caption[Muon Detector Spectrum]{Gamma-ray and muon energy spectra measured via a single muon panel in surface calibration run. The blue and red curves represent the fitted gamma-ray and muon contributions, respectively, with the combined fit shown in cyan. The orange distribution corresponds to muon events when applying a four-panel coincidence requirement. The dashed green line is the energy threshold applied to select muons.}
        \label{fig:muonSpec}
    \end{figure}

    Muon events were identified by a quadruple coincidence requirement within a 40\,ns acceptance window along with an energy threshold of 3.6\,MeV in three adjacent panels\footnote{Allowing lower energy depositions in either the top or bottom panel increased geometric efficiency by including possible ``corner-clipping'' events.}. This threshold was determined via a Landau-Gaussian fit to the surface data and placed to maximise muon acceptance as depicted in Fig.~\ref{fig:muonSpec}. The goodness of this threshold is demonstrated as all quadruple coincidence events within the middle two panels were above this threshold. The muon detection efficiency was determined to be 29.6~\textpm~2.0\,\% based on the expected surface flux of $1.67\times10^{-2} \mu/\textrm{cm}/\textrm{s}$ \cite{ramesh2012fluxvariationcosmicmuons}.

    \subsection{Flux Measurement}

    The muon measurement campaign at the TM3500 site in Bedretto lasted 21 days. During this time, 8 muon events were recorded in the signal region defined by data selections, yielding a measured underground muon flux of $(2.54~\pm~0.97)~\times 10^{-8}~\mu/\mathrm{s}/\mathrm{cm}^2$. Uncertainty contributions include statistical (35.4~\%), variance in detection rate (6.6~\%), and uncertainty in the energy scale for the selection requirements (12.5~\%).  

    This measured flux was verified utilising Monte-Carlo simulation frameworks MUon SImulation Code (MUSIC) \cite{music} and MUon Simulations UNderground (MUSUN) \cite{musun} along with a Geant4 simulation of the telescope. A water depth equivalent of 3998\,m.w.e. was estimated based on topographical data provided by the Swiss Federal Office of Topography~\cite{swisstopo}.  The measured and simulated fluxes are in good agreement. Presented in Fig.~\ref{fig:waterdepth} is a comparison to other underground laboratories. Notably, the muon flux at the Bedretto lab is between that of Laboratori Nazionali del Gran Sasso (LNGS) and the Laboratoire Souterrain de Modane (LSM) and is therefore competitive on the global stage.

\section{Gamma Ray Flux}
\label{sec:gamma_flux}
To assess the natural radioactivity of the granite wall and inform Monte-Carlo-based predictions of gamma and neutron fluxes, we measured different granite samples with the Gator high-purity germanium detector (HPGe) facility at LNGS. 

\subsection{HPGe Screening of Granite Samples}
    
\begin{figure*}[t]
        \centering
        \includegraphics[width=\textwidth]{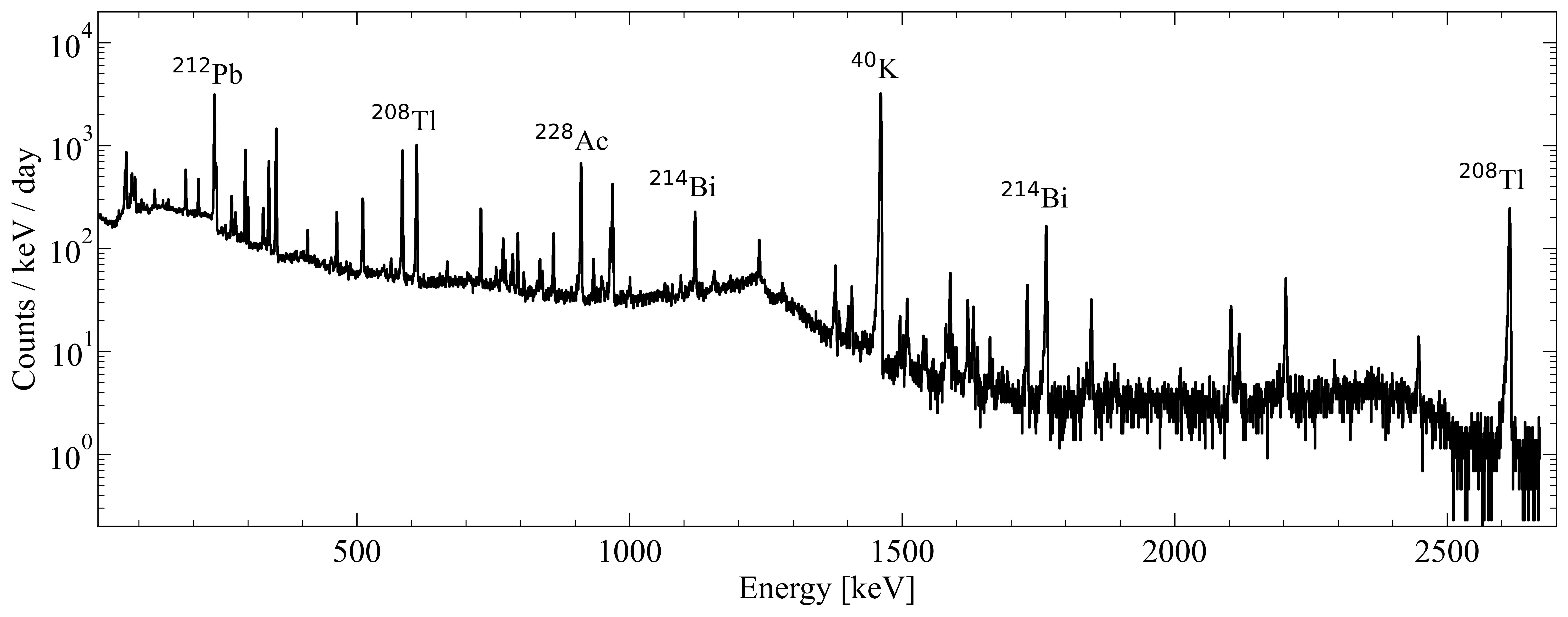}
        \caption{Calibrated energy spectrum from the TM3500 Bedretto granite measured in Gator. The spectra are dominated by gamma lines of the $^{238}$U and $^{232}$Th decay chains; the $^{40}$K line at 1460\,keV is also seen.} 
        \label{fig:gamma_rock}
    \end{figure*}

Gator is a dedicated counting station consisting of a 2.2\,kg p-type Ge crystal inside a low-background copper cryostat and shield~\cite{Araujo:2022kip}.
The detector background rate is $82.0 \pm 0.7$ counts/(d$\cdot$kg) in the 100–2700\,keV range, enabling sensitivities at the $\mu$Bq/kg level~\cite{Baudis:2011am,Araujo:2022kip}. Two granite rock samples were taken from the Bedretto tunnel for gamma counting. A 38\,g cube with a length of 2.5\,cm from TM3500 as well as a cylindrical core sample taken from TM2500. The core sample was machined into two cylinders with a diameter 82~mm and heights 49.5~mm and 52.0~mm respectively. The TM3500 sample was screened in the Gator facility for 5\,days, the TM2500 sample was counted for approximately 23~hours. The measured gamma-ray spectrum from the TM3500 sample is shown in Fig.~\ref{fig:gamma_rock}. The activities of the major decay chains for the TM3500 and TM2500 rock samples are listed in Table~\ref{tab:rock_activity_final}. The sample from TM3500 possesses an about $35$\,\% lower activity than the sample from T2500 as consistent with the gamma flux measurement shown in Fig.~\ref{fig:tunnel_gamma}

\subsection{Equilibrium of the TM2500 Sample}
    
    The analysis of the $^{238}$U decay chain via the larger TM2500 sample was found to be out of secular equilibrium. The early chain activity was determined via the 1001\,keV gamma ray from the decay of $^{234}$Pa. The entire $^{238}$U decay chain activity as determined by this single decay would be 220.6~\textpm~18.0\,Bq/kg. The late chain activity was determined utilizing the weighted average of $^{214}$Bi (609.3, 1120.3, 1764.5\,keV) and $^{214}$Pb (295.2\,keV and 351.9\,keV) gamma rays. From these lines the chain activity would be 161.4\,Bq/kg, $\simeq 27 \%$ lower than the activity computed from the decay of $^{234}$Pa. 

    From this discrepancy, we hypothesise the $^{238}$U chain to be out of secular equilibrium with the equilibrium broken at $^{226}$Ra due to the solubility of Ra, leaching out of the rock. This hypothesis was independently confirmed by de-convolving the 185.7\,keV gamma~ray of $^{235}$U decay from the 186.2\,keV gamma~ray of $^{226}$Ra. The $^{238}$U decay chain activity as calculated from the $^{235}$U peak was found to be 216~\textpm~12~Bq/kg, in good agreement with the early chain activity 220.6~\textpm~16\,Bq/kg as calculated from $^{234}$Pa\footnote{For the early chain TM3500 sample, the equilibrium is assumed to have the same break as the TM2500 sample.}. 

    The $^{232}$Th chain activity was found to be in secular equilibrium and was determined from averaging the activities of $^{208}$Tl, $^{228}$Ac, and $^{212}$Pb decays. The chain activity was found to be 118~\textpm~3\,Bq/kg. Although the $^{235}$U chain contains $^{223}$Ra, it's half-life is only 11.43\,days, opposed to 1600~years for $^{226}$Ra. We therefore assume the $^{223}$Ra decays before it is able to leach out of the rock and therefore the $^{235}$U decay chain is in secular equilibrium.

\begin{table}
\centering
\caption{Activities of the major isotopic decay chains of the granite samples taken from TM3500 and TM2500 as determined by the Gator counting facility.}
\label{tab:rock_activity_final}
\begin{tabular}{c D{,}{\pm}{-1} D{,}{\pm}{-1}}
\hline\noalign{\smallskip}
\multirow{1}{*}{\textbf{Isotope}} & \multicolumn{2}{c}{\textbf{Activity [Bq/kg]}} \\
\noalign{\smallskip}\hline
                                  &     \multicolumn{1}{c}{TM3500}   &   \multicolumn{1}{c}{TM2500} \\
\noalign{\smallskip}\cline{2-3}\noalign{\smallskip}
$^{235}$U Chain                   &                1.8\,,\,0.1       & 10.0\,,\,0.6  \\
$^{232}$Th Chain                  &                83.8\,,\,1.9       & 118\,,\,3  \\
$^{238}$U -  $^{230}$Th           &                77.4\,,\,6.8   & 221\,,\,18 \\
$^{226}$Ra - $^{206}$Pb           &                56.6\,,\, 1.3       & 161\,,\,4 \\ 
$^{40}$K                          &                1090\,,\,50       & 1130\,,\,70 \\
\noalign{\smallskip}\hline
\end{tabular}
\end{table}

\subsection{Gamma Flux Measurement}

The gamma-ray flux was measured and mapped within the Bedretto tunnel utilizing a $3 \times 3$~in NaI[Tl] gamma-ray detector. A large variation is observed strongly correlated with the local geological rock type as illustrated in Fig.~\ref{fig:tunnel_gamma}. A 35~\% decrease in measured rate is observed between the porphyritic and rotondo granite. A local minimum in the flux is observed between approx. 3000 and 3300\,m. The flux from the prominent high energy lines from the decays of $^{40}$K, $^{214}$Bi, and $^{208}$Tl is measured as $3.72 \pm 0.28, 0.95 \pm 0.16$, and $0.95 \pm 0.19~\gamma/\mathrm{cm}^2/\mathrm{sec}$. Typically underground labs walls are reinforced with shotcrete for shielding and safety reasons. 
A 15\,cm thick wall would attenuate the hard gamma flux by approximately 90~\%~\cite{ConcreteShielding}, resulting in levels comparable to the total gamma ray fluxes measured at LNGS $0.3~\gamma/\mathrm{cm}^2/\mathrm{sec}$~\cite{Haffke:2011fp}) and  $0.6~\gamma/\mathrm{cm}^2/\mathrm{sec}$~\cite{LSMgamma} at LSM after laboratory outfitting.

\section{Neutron Flux}
    
    Measurements of the neutron flux were performed using a \textit{LND 252108} $^{3}$He proportional counter with an active volume of 89.01~cm$^{3}$ and a fill pressure of 7600~Torr~\cite{LND}. The thermal neutron sensitivity is 34.0~counts per second (CPS) per neutron/cm$^{2}$/s~\cite{LND} which was verified by simulation. The tube was biased at 2~kV utilizing a CAEN DT5533E~\cite{CAEN5533E} power supply. Pulses were formed and read-out utilizing an Ortec 142PC preamplifier and a CAEN DT5724 100~MHz digitizer~\cite{Ortec,CAEN5724}.

\begin{figure}[h]
    \centering
    \includegraphics[width=0.45\textwidth]{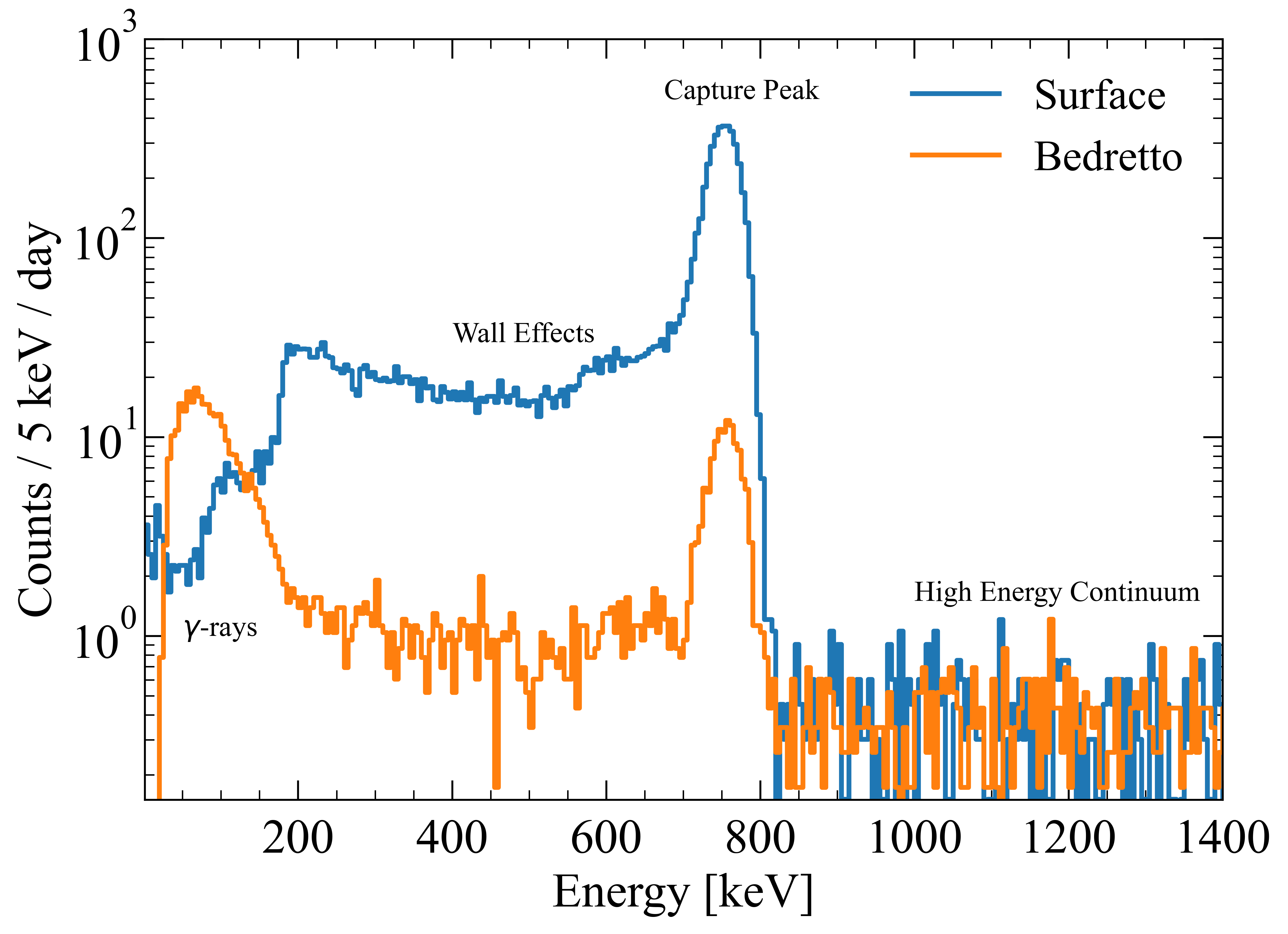}
    \caption{$^{3}$He neutron spectrum of the ambient neutron flux as measured on the surface and at TM3500. The primary neutron capture peak is visible at 765~keV. The low energy structure is from gamma and X-ray interactions with the detector.}\label{fig:He3Spec}
    \end{figure}

    The typical spectral response of $^{3}$He proportional counters to thermal neutron irradiation comprise a prominent capture peak at 765~keV. This corresponds to the Q-value of the $(n,p)$ reaction on $^{3}$He. Wall effects extend below the capture peak with steps at 574~keV and 191~keV as seen in Fig.~\ref{fig:He3Spec}. The ambient surface thermal neutron flux is measured to be $(2.6 \pm 0.1)\times10^{-3}~\mathrm{neutrons}/\mathrm{cm}^2/\mathrm{sec}$ which is consistent with accepted values of the earth's neutron albedo~\cite{korun_martincic_pucelj_ravnik_1996,Gorshkov1964}. Other spectral features include a high energy continuum above the neutron capture peak. This typically arises from alpha contamination within the tube wall~\cite{HASHEMINEZHAD1998100}. At low energies, below 191~keV, are double wall effects and interactions from x-rays and gamma-rays, the energy deposition from which are limited by the mean free path of electrons within the gaseous detector volume.

    The measurement campaign lasted 21~days, with the $^{3}$He counter at TM3500 in two configurations: the unmoderated tube and with a Polyoxymethylene (POM) plastic neutron moderator. The plastic moderator is a cylinder with diameter of 13.5~cm and length of 30~cm. A 27~cm aperture into the moderator allows to completely enclose the $^{3}$He counter. 

    \subsection{Thermal Neutron Flux}
    
    The neutron capture rate was calculated with the area of the 765~keV Q-energy peak. The fraction of thermal neutron events within the peak was found to be 59.7~\textpm~2.8~\% which was determined using surface data by calculating the ratio of peak events to wall effects after subtracting the alpha continuum. This estimate is consistent with the values determined for other $^{3}$He counters of the same diameter~\cite{debickiLNGS}.

    The area of the 765~keV capture peak was determined by simultaneously fitting an exponentially modified Gaussian on top a resolution-smeared step-function background model. For the unmoderated tube, the peak area is found to be 1125~$\pm$~37 counts with a capture rate of $(1.13 \pm 0.04)\times 10^{-2}$~CPS. Sources of uncertainty include event selections, peak fitting, and choice of fit model. The thermal neutron flux at TM3500 is therefore estimated to be $(5.56 \pm 0.26)\times10^{-5}~\mathrm{neutrons}/\mathrm{cm}^2/\mathrm{sec}$. This flux was independently verified utilizing an enriched $^{6}$LiI(Eu) neutron detector~\cite{Tarka_2012}.

    The low energy peak observed below about $200$~keV in the Bedretto data are attributed to X-ray and gamma-ray scattering interactions. We were able to reproduce the excess when placing a natural thorium source next to the detector during surface testing. The response of $^{3}$He proportional counters has been studied as a function of gamma-ray flux and shows similar low energy structures such as in Ref.~\cite{CHOI2019263}. Similar as for gamma-ray flux, the measured neutron flux of the bare rock walls is as expected higher than the one in outfitted underground laboratories such as LSM or LNGS where the thermal neutron fluxes are and approximately $(1.1 \pm 0.1) \times10^{-6}~ \mathrm{neutrons}/\mathrm{cm}^2/\mathrm{sec}$ and $(0.38 \pm 0.14) \times 10^{-6}$ \\ $\mathrm{neutrons}/\mathrm{cm}^2/\mathrm{sec}$ \cite{RLemrani_2006,Wulandari:2003cr,Ascenzo:2025ujf,LNGSfast} due to the exposed rock wall. Outfitting the laboratory, potentially adding neutron moderators such as boron to the shotcrete, will attenuate the neutron flux to comparable levels or lower as in other laboratories.

    \subsection{Fast Neutron Flux}
    In the case with the moderator, the capture rate was found to be $(8.74 \pm 0.24)\times10^{-3}$ captures per second within the 765~keV peak. The fast neutron flux is extracted utilizing MCNPX-Polimi simulations~\cite{MCNPX,MCNPX-Polimi-1,CLARKE2013135}. The thermal neutron reflectance and absorption from the POM moderator was simulated to be 57~\textpm~5~\%. The fast neutron flux was assumed to be spectrally similar to that of Laboratoire souterrain de Modane (LSM) as reported in Reference~\cite{RLemrani_2006}. Based on this assumption we estimate a fast neutron flux of $(6.3 \pm 0.5)\times 10^{-5}~\mathrm{neutrons}/\mathrm{cm}^2/\mathrm{sec}$.  This flux however relies on a fundamental assumption that the neutron energy spectrum in similar to that at LSM. To more accurately measure the fast neutron flux, we plan a follow up campaign utilizing both fast neutron detectors and the capture gated neutron spectroscopy technique similar to that discussed in Ref.~\cite{CaptureGatedNeutron}.

\section{Radon Concentration}
    The uranium and thorium contained in the rock, see also Fig.~\ref{sec:gamma_flux}, continuously produce radon which emanates from the rock into the cavern air. In absence of natural and/or forced airflow, the radon activity concentration in the tunnel air equilibrates with the radon released from the rock.
    However, air handling infrastructure, continuously delivering fresh air from the outside into the cavern through a duct terminating at about TM2500, dictates the concentration and distribution of radon in the cavern. The average ventilation rate is maintained at one tunnel volume per about $2.5$\,h, yielding an average cavern air current of about $0.3$\,m/s. 

    The radon concentration at several points inside the tunnel was measured using a commercial electrostatic RAD7 radon monitor from Durridge \cite{rad7}. To prevent degradation of the detection efficiency, the air was dehumidified using a laboratory gas drying unit filled with about 0.5\,kg Drierite desiccant. Measurements of $^{222}$Rn concentration near the intake and directly next to the output of the air handling duct at about TM2500 yield near identical concentrations, in the range of 1-40~Bq/m$^3$, depending on environmental conditions. Between TM2500 and the entrance, the continuous air flow results in an increasing gradient in radon concentration with  $300 \pm 50$~Bq/m$^3$ and $1200 \pm 200$~Bq/m$^3$ at TM1500 and TM500 respectively. Measurements deeper than the air-handling output, where forced air circulation is expected to be minimal, show a radon concentration of 2-4~kBq/m$^3$. Measurements of $^{220}$Rn concentration were only briefly pursued using a reduced desiccant mass of 30\,g. They confirmed a sub-dominant contribution due to the much shorter half-life of this isotope.
    
    A second RAD7 detector was deployed at TM3500 for long-term monitoring of radon concentration. Desiccant consumption rate was reduced using a Durridge Drystik humidity exchanger\,\cite{drystik}. Radon concentration was measured in 30 minutes intervals between the end of July and middle of December 2025 with few interruptions for maintenance. The data revealed significant and periodic fluctuations of the radon concentration in the range of 1.5-3\,kBq/m$^3$, with fluctuations of up to 4.5\,kBq/m$^3$, see also Fig.~\ref{fig:radon_time_evolution}. Causes for the observed variation are assumed to be related to forced air flow and pressure changes induced by the train schedule in the Furka tunnel.

    Other underground labs feature radon concentration varying between 2\,Bq/m$^3$\,\cite{Scovell:2023pxo} and 300\,Bq/m$^3$\,\cite{Heise:2022iaf} for Boulby and the SURF underground laboratory respectively. LNGS, with about 20\,Bq/m$^3$\,\cite{ampollini2023sub}, is the closest in observed radon concentration, likely due to the very similar rock composition of the regions. The present measurements underline that the radon concentration can be sufficiently suppressed by a continuous and targeted supply of outside-air.

\section{Magnetic Background}

The magnetic field background at TM3500 was measured using a \textit{Bartington Mag-13MS70} three-axis fluxgate magnetometer (powered by a \textit{PSU-1} power supply) for a total duration of \SI{471.8}{\hour}~\cite{bartington_instruments_mag-13_2024,bartington_instruments_magnetometer_2024}.
The combined sensor and power supply noise floor is $\leq\SI{7}{\pico\tesla\per\sqrt{\hertz}}$ at \SI{1}{\hertz}.
The data acquisition was carried out using a \textit{Digilent MCC128} DAQ hat with a \SI{16}{bit} ADC resolution and an equivalent noise of $2\rm nT_{rms}$~\cite{measurement_computing_corporation_16-bit_202}.
The sampling rate was set to $20\rm kHz$ per axis to avoid aliasing given the \SI{9.5}{\kilo\hertz} low pass filter of the \textit{PSU-1}.
At $f_s = 20\rm kHz$ the quantization noise from the DAQ is estimated to be $\sim \frac{\Delta}{\sqrt{6f_s}}=5\rm pT/\sqrt{Hz}$.
To fully characterise the combined noise floor of the measurement setup and compare to the actual field background a 3-layer MuMetal zero-gauss chamber with an attenuation of \SI{>1000}{} was used to perform an on-off test~\cite{magnetic_shield_corp_zg-3_2013}.
Of the 471.8 hours of measurement, 161.6 hours were taken with the magnetometer inside the zero-gauss chamber and 310.2 hours were taken without the zero-gauss chamber.

The one-sided Power spectral density (PSD) of the magnetic field background is normalised such that:
$$
\int_0^{f_s/2}\mathrm{PSD_{avg}}(f) \mathrm{d}f = B_{\rm avg}^2,
$$
with $B_{\rm avg}^2 = \frac{1}{3} \left(B_{x}^2+B_{y}^2+B_{z}^2\right)$ being the squared average across three directions. We estimate the power spectral density of the magnetic noise floor using Welsch's method with a hamming averaging window o $100\rm{s}$~\cite{Heinzeldft}.

\begin{figure}[h]
    \centering
    \includegraphics[width=1\linewidth]{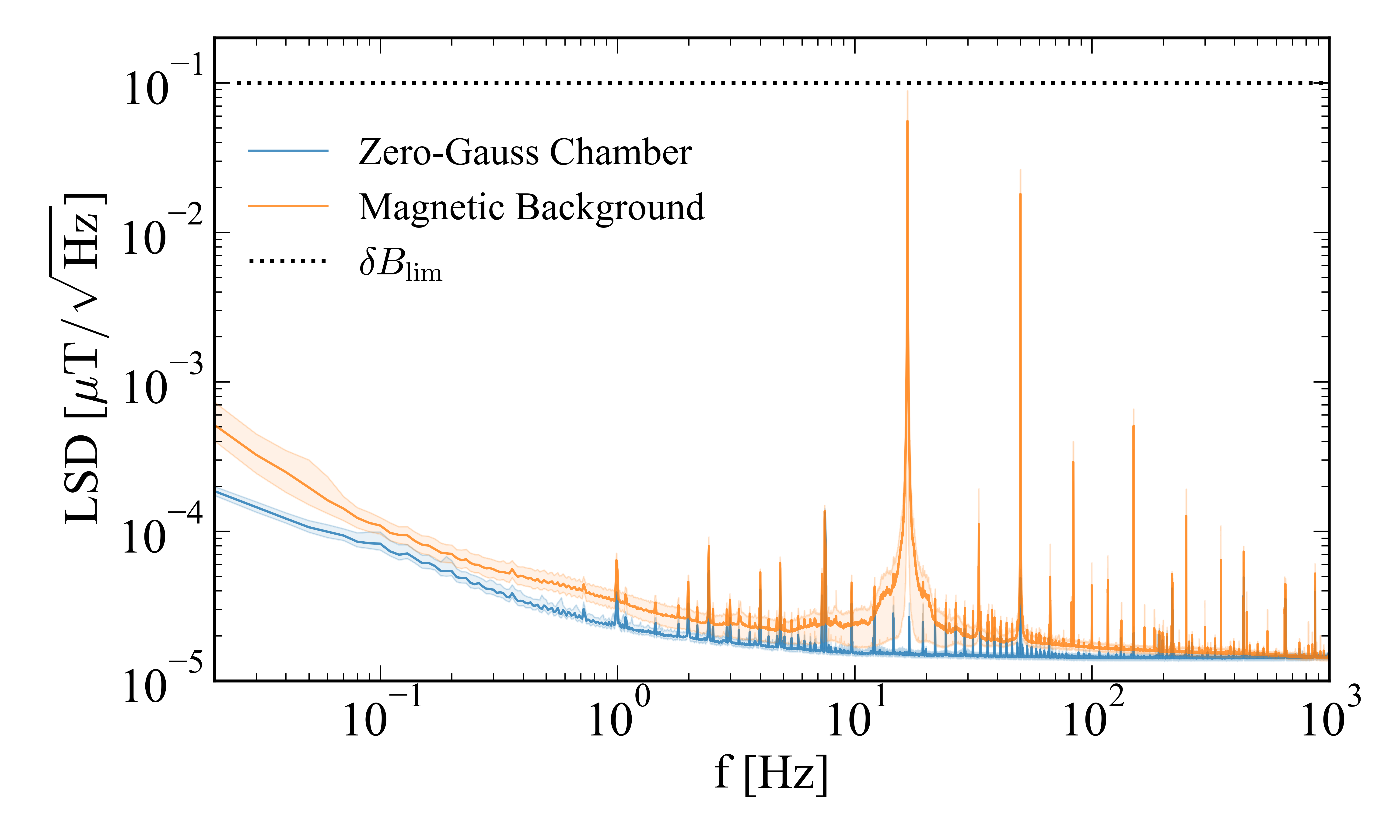}
    \caption{The linear spectral density (LSD) of magnetic field noise floor magnitude, which is defined as the square root of the power spectral density (PSD). We show the noise floor measured inside zero-gauss chamber (blue), and without zero-gauss chamber (orange). The filled ranges denote the 16th and 84th percentiles of the distribution of per-hour LSD values, and the solid lines denote the median of the LSDs.}
    \label{fig:magpsd}
\end{figure}

Figure~\ref{fig:magpsd} shows the comparison of the zero-gauss chamber on-off test.
In the frequency range of $2\times10^{-2}-10^{2}~\rm{Hz}$ the noise floor of the setup, as measured with the zero-gauss chamber, is significantly separated from the magnetic field measured without the zero-gauss chamber.
This confirms that our setup is sensitive enough to record the effective magnetic field background.

The magnetic field PSD ankles at \SI{\sim 1}{\hertz} with a white noise floor at $\sim$\SI{15}{\pico \tesla \per \sqrt{\hertz}}.
This is lower than the $100\rm nT/\sqrt{Hz}$ requirement for building Strontium-based long baseline atomic interferometer gravitational wave (AI-GW) detectors and for other potential applications, as discussed in Ref.~\cite{MAGIS-100-2021etm}.
However, there is a significant peak centered on \SI{16.7}{\hertz}, which corresponds to the nominal frequency of the alternating current supply used by the Furka railway~\cite{jossen_orion-triebzuge_2024}.
Analyzing a subset of the data between the hours 22:00 and 04:00 shows a strong reduction of this peak, which is well correlated with the absence of regular train operation during these hours~\cite{systemaufgaben_kundeninformation_fahrplanfeld_2024,systemaufgaben_kundeninformation_fahrplanfeld_2024-1}.

Similar measurements have been carried out at CERN, in the context of feasibility studies for AICE~\cite{arduiniLongBaselineAtomInterferometer2023}. Although, those measurements are dominated by their setup noise floor, comparison with our result shows that Bedretto's magnetic field background - with the exception of the peak at \SI{16.7}{\hertz} - is very competitive. The zero-gauss chamber measurements show that such magnetic field contaminations can be reduced effectively with appropriate shielding.

\section{Seismic Background}
    
\begin{figure}
    \centering
    \includegraphics[width=1\linewidth]{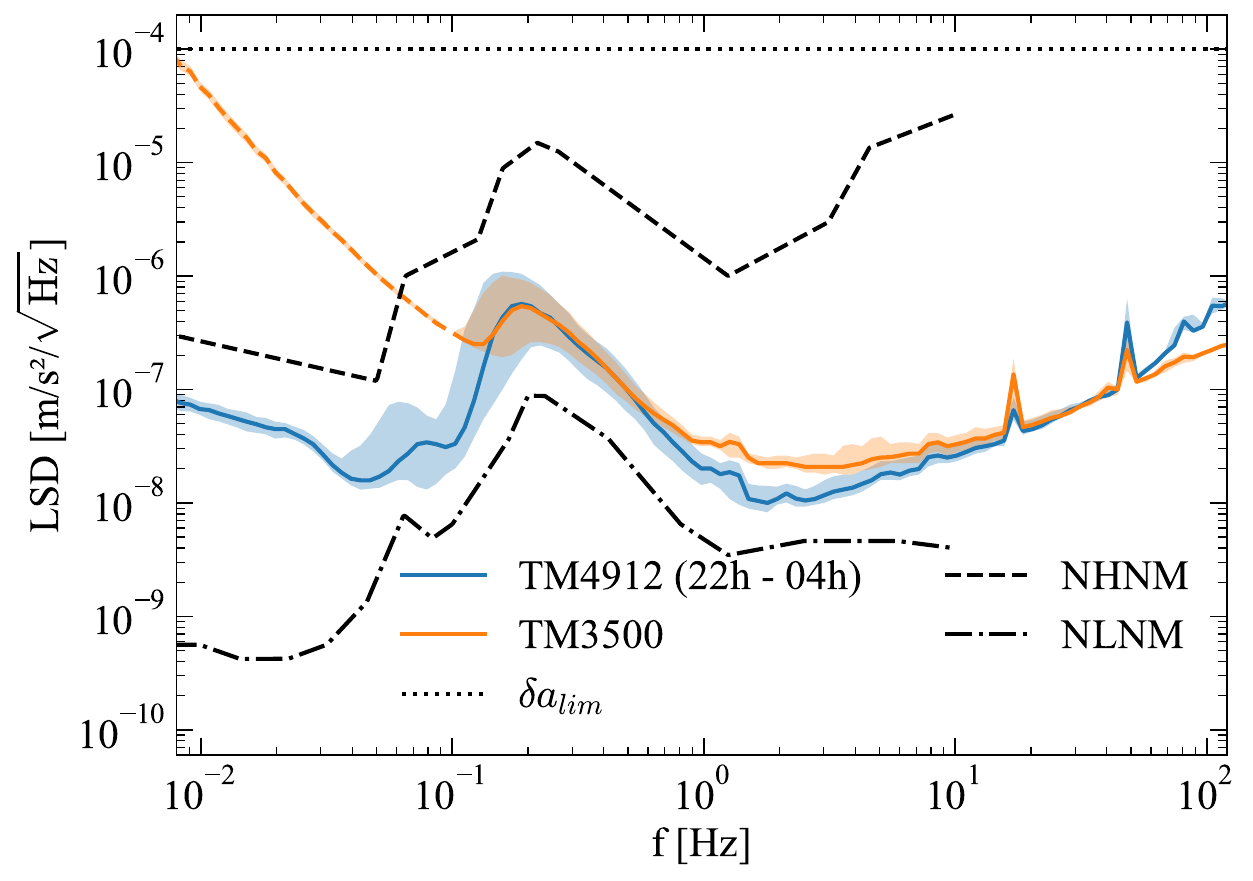}
    \caption{The linear spectral density (LSD) of the measured seismic activity.
    The orange curve shows the seismic noise at TM3500.
    This measurement is limited by sensor noise below \SI{1e-1}{\hertz}.
    The night time seismic noise measured at TM4912 is shown as a best case reference.
    The filled ranges denote the 16th and 84th percentiles of the distribution of per \SI{30}{\minute} LSD values, and the solid lines denote the median of the LSDs.}
    \label{fig:seismic_psd}
\end{figure}

Multiple seismometers are permanently installed in the Bedretto tunnel, as part of the measurement network of the Swiss Seismological Service~\cite{swiss_seismological_service_swiss_2016}.
At TM3500 a short period \textit{Lennartz LE-3DliteMkIII} seismometer with a corner period of \SI{1}{\second} is installed, sampled at \SI{500}{sps}.
The average seismic acceleration across the three directions measured with this sensor is shown orange in Fig.~\ref{fig:seismic_psd}.
As for the magnetic field measurements, the data is presented in the form of a linear spectral density, defined as the square root of the power spectral density.
Individual LSD curves are calculated per \SI{30}{\minute} interval, and the distribution of these LSD curves is indicated with the shaded areas in Fig.~\ref{fig:seismic_psd} (16th to 84th percentile region).
The data covers a one year period from October 2024 to October 2025.
Also indicated are the LSD corresponding to the New High and New Low Noise Models (NHNM and NLNM)~\cite{peterson_observations_1993}.
Two small peaks are visible at higher frequencies, one at \SI{16.7}{\hertz} and one at \SI{50}{\hertz}, corresponding to the overhead line frequency of the Furka railway~\cite{jossen_orion-triebzuge_2024} and the frequency of the Swiss power grid respectively. 

Below \SI{1e-1}{\hertz} the measurements at TM3500 are dominated by the noise floor of the short period seismometer.
As a best case reference the data from a broadband \textit{Nanometrics Trillium Compact} seismometer with a \SI{120}{\second} corner period, installed at TM4912 is also shown in Fig.~\ref{fig:seismic_psd} (blue curve).
This sensor is located very close to the Furka railway tunnel, and thus only data from night times (22:00 to 04:00) is shown, when no regular trains are running~\cite{systemaufgaben_kundeninformation_fahrplanfeld_2024,systemaufgaben_kundeninformation_fahrplanfeld_2024-1}.
Below \SI{2.5e-2}{\hertz} the measurement at TM4912 is again dominated by the sensor noise floor.

In order to limit the noise contributions from vibration of the optical components in AI-GW detectors, the seismic acceleration spectrum should be below $\delta{}a_{lim} = \SI{1e-4}{\meter\per\second\squared\per\sqrt{Hz}}$~\cite{MAGIS-100-2021etm,arduiniLongBaselineAtomInterferometer2023}.
The Bedretto site exceeds this criteria by two orders of magnitude.

Overall, the seismic activity at TM3500 is very low, clearly below the NHNM, important to limit the contributions of gravity gradient noise to AI-GW detectors~\cite{MAGIS-100-2021etm}.
Comparing the seismic activity at Bedretto with similar measurements carried out at CERN (for AICE~\cite{arduiniLongBaselineAtomInterferometer2023}) and at Fermilab (for MAGIS-100~\cite{mitchellMAGIS100EnvironmentalCharacterization2022}) shows the extremely low seismic background of the Bedretto site.
This is especially the case in the range \SIrange{1}{100}{\hertz}, where seismic noise is highly dominated by human activity and where the Bedretto site shows a seismic noise level of at least one order of magnitude lower compared to the other sites.

\section{Discussion}
    The comprehensive  measurement of the background environment at the BedrettoLab demonstrates the suitability of the site for a large variety of underground experiments and provides the data necessary to plan future fundamental physics projects. The planned Bedretto Underground Lab for Fundamental Physics at TM3500 will be one of the deepest underground sites in Europe, exceptionally quiet in terms of vibrations and E\&M shielding. After standard shielding it is expected to achieve similar or better levels of ionizing backgrounds as comparable laboratories. The comprehensive assessment of background levels allows for optimised construction, such as using borated shotcrete to further reduce neutron levels.
    Furthermore, the long-term monitoring of backgrounds will characterise seasonal cycles and further establish diurnal or weekly patterns, for example E\&M interference,  and transient events such as human activities. The high-resolution instrumentation of the rock with geological sensors allows for very detailed understanding of vibrational backgrounds.  Finally, the location and nature of the site allows for excellent accessibility and the easy excavation of new caverns and expansions.
    

    Underground laboratories necessarily cover bare rock walls with shotcrete. A standard 15~cm concrete walls will reduce the thermal neutron flux by a factor of 10 and the fast neutron flux by 85~\% as simulated with MCNP which is consistent with the reduction factor shown in Ref.~\cite{ConcreteShielding}. 
    This protective layer of concrete will also reduce the ambient gamma flux, for the dominant 1.4~MeV $^{40}$K gamma-ray, a 15~cm thick concrete layer will attenuate the flux by approximately 90~\%, while standard paint-on radon barriers will mitigate radon emanation.  With these standard  measures in place, the resulting neutron and gamma backgrounds are expected to be comparable or lower to those reported for the established European underground laboratories.

    The site allows for easy and fast excavations and expansion of existing underground space. Our plan is to developed a sizeable O(100 m$^2$) underground clean laboratory for underground R\&D for rare event searches, quantum sensing, gravitational waves and others quickly. Future expansion are possible including excavation of O(100m-1km) tunnels and shafts for, e.g., atom interferometry, gravitational wave and neutrino detectors. The site offers excellent accessibility to leading research institutions and major cities such as Z\"urich, Geneva, and Milano. Additional logistical benefits are the horizontal access, the capacity to transport up to 18\,t of mass at once and existing surface infrastructure.

\section{Conclusion}
We performed a comprehensive set of background measurements in the bare rock wall environment of the Bedretto tunnel in Ticino, Switzerland. These measurements, combined with simulations of  shot\-crete shield\-ing, demonstrates the suitability of the site for a large variety of underground experiments and provides the data necessary to plan future fundamental physics projects. With the installation of shotcrete lining and a ventilation system, the site will rank among the deepest and quietest locations globally. These preparatory measurements also allow for optimised infrastructure design when furnishing the laboratory, allowing us to construct a world-leading facility with horizontal access and proximity to national and international metropolitan centres. 
Initially, the laboratory will serve as a low-background R\&D facility with close ties to leading Swiss institutions such as the University of Z\"urich, University of Geneva, University of Bern and EHT Z\"urich. Operating existing experimental setups for rare-event searches, gravitational wave physics, quantum sensing, HPGe counters and more underground will lead to significant sensitivity improvements and greatly advancing existing and planned research.  In the long-term the site can be very suitable also to host major international efforts due to the ability to quickly excavate and prepare additional laboratory space.

\begin{acknowledgements}
The research for this work took place at the BedrettoLab in Bedretto, Ticiono, Switzerland. Funding for this work is provided by the University of
Zurich. We are grateful to the assistance of the Department of Earth and Planetary Sciences at ETH Z\"urich and its personnel, in particular BedrettoLab,
in providing physical access and general logistical and technical support. We extend our gratitude to Marco Bertoldi for his generous hospitality, support and the insightful discussions held during our visits. We also want to thank the the Swiss Seismological Service and Swiss Topo for use of their data. Philippe Jetzer thanks the Tomalla Foundation for partial support.
\end{acknowledgements}

\bibliographystyle{JHEP}
\bibliography{sn-bibliography}

@article{Yemilab_Muon,
    author = {Y. Kim and H.S. Lee},
    title = "{Yemilab, a new underground laboratory in Korea}",
    journal = {AAPPS Bulletin},
    year = {2024},
    url = {https://doi.org/10.1007/s43673-024-00132-8},
    doi = {10.1007/s43673-024-00132-8}
}

@article{Soudan_Muon,
    author = "Zhang, C. and Mei, D. -M.",
    title = "{Measuring Muon-Induced Neutrons with Liquid Scintillation Detector at Soudan Mine}",
    eprint = "1407.3246",
    archivePrefix = "arXiv",
    primaryClass = "physics.ins-det",
    doi = "10.1103/PhysRevD.90.122003",
    journal = "Phys. Rev. D",
    volume = "90",
    number = "12",
    pages = "122003",
    year = "2014"
}

@article{Boulby_Muon,
    author = "Robinson, M. and others",
    title = "{Measurements of muon flux at 1070 meters vertical depth in the Boulby underground laboratory}",
    eprint = "hep-ex/0306014",
    archivePrefix = "arXiv",
    doi = "10.1016/S0168-9002(03)01973-9",
    journal = "Nucl. Instrum. Meth. A",
    volume = "511",
    pages = "347--353",
    year = "2003"
}

@article{LSC_Muon,
    author = "Trzaska, Wladyslaw Henryk and others",
    title = "{Cosmic-ray muon flux at Canfranc Underground Laboratory}",
    eprint = "1902.00868",
    archivePrefix = "arXiv",
    primaryClass = "physics.ins-det",
    doi = "10.1140/epjc/s10052-019-7239-9",
    journal = "Eur. Phys. J. C",
    volume = "79",
    number = "8",
    pages = "721",
    year = "2019"
}

@article{LNGS_Muon,
    author = "Agostini, M. and others",
    collaboration = "Borexino",
    title = "{Modulations of the Cosmic Muon Signal in Ten Years of Borexino Data}",
    eprint = "1808.04207",
    archivePrefix = "arXiv",
    primaryClass = "hep-ex",
    doi = "10.1088/1475-7516/2019/02/046",
    journal = "JCAP",
    volume = "02",
    pages = "046",
    year = "2019"
}

@article{LSM_Muon,
    author = "Schmidt, B. and others",
    collaboration = "EDELWEISS",
    title = "{Muon-induced background in the EDELWEISS dark matter search}",
    eprint = "1302.7112",
    archivePrefix = "arXiv",
    primaryClass = "astro-ph.CO",
    doi = "10.1016/j.astropartphys.2013.01.014",
    journal = "Astropart. Phys.",
    volume = "44",
    pages = "28--39",
    year = "2013"
}

@article{SURF_Muon,
    author = "Abgrall, N. and others",
    collaboration = "MAJORANA",
    title = "{Muon Flux Measurements at the Davis Campus of the Sanford Underground Research Facility with the Majorana Demonstrator Veto System}",
    eprint = "1602.07742",
    archivePrefix = "arXiv",
    primaryClass = "nucl-ex",
    doi = "10.1016/j.astropartphys.2017.01.013",
    journal = "Astropart. Phys.",
    volume = "93",
    pages = "70--75",
    year = "2017"
}

@article{CJPL_Muon,
    author = "Guo, Ziyi and others",
    collaboration = "JNE",
    title = "{Muon flux measurement at China Jinping Underground Laboratory}",
    eprint = "2007.15925",
    archivePrefix = "arXiv",
    primaryClass = "physics.ins-det",
    doi = "10.1088/1674-1137/abccae",
    journal = "Chin. Phys. C",
    volume = "45",
    number = "2",
    pages = "025001",
    year = "2021"
}

@article{SNO_Muon,
    author = "Aharmim, B. and others",
    collaboration = "SNO",
    title = "{Measurement of the Cosmic Ray and Neutrino-Induced Muon Flux at the Sudbury Neutrino Observatory}",
    eprint = "0902.2776",
    archivePrefix = "arXiv",
    primaryClass = "hep-ex",
    doi = "10.1103/PhysRevD.80.012001",
    journal = "Phys. Rev. D",
    volume = "80",
    pages = "012001",
    year = "2009"
}

@article{Woodley:2024eln,
    author = "Woodley, William and Fedynitch, Anatoli and Piro, Marie-C{\'e}cile",
    title = "{Cosmic ray muons in laboratories deep underground}",
    eprint = "2406.10339",
    archivePrefix = "arXiv",
    primaryClass = "hep-ph",
    doi = "10.1103/PhysRevD.110.063006",
    journal = "Phys. Rev. D",
    volume = "110",
    number = "6",
    pages = "063006",
    year = "2024"
}

@article{music,
    author = "Antonioli, P. and Ghetti, C. and Korolkova, E. V. and Kudryavtsev, V. A. and Sartorelli, G.",
    title = "{A Three-dimensional code for muon propagation through the rock: Music}",
    eprint = "hep-ph/9705408",
    archivePrefix = "arXiv",
    reportNumber = "INFN-AE-97-12",
    doi = "10.1016/S0927-6505(97)00035-2",
    journal = "Astropart. Phys.",
    volume = "7",
    pages = "357--368",
    year = "1997"
}

@article{ramesh2012fluxvariationcosmicmuons,
    author = "Ramesh, Nepal and Hawron, Martin and Martin, Clayton and Bachri, Abdel",
    title = "{Flux Variation of Cosmic Muons}",
    eprint = "1203.0101",
    archivePrefix = "arXiv",
    primaryClass = "physics.ins-det",
    journal = "J. Arkansas Acad. Sci.",
    volume = "65",
    pages = "67--72",
    year = "2011"
}

@article{musun, 
    author = "Kudryavtsev, V. A.",
    title = "{Muon simulation codes MUSIC and MUSUN for underground physics}",
    eprint = "0810.4635",
    archivePrefix = "arXiv",
    primaryClass = "physics.comp-ph",
    doi = "10.1016/j.cpc.2008.10.013",
    journal = "Comput. Phys. Commun.",
    volume = "180",
    pages = "339--346",
    year = "2009"
}

@article{Bruno:2019vad,
    author = "Bruno, G. and Fulgione, W.",
    title = "{Flux measurement of fast neutrons in the Gran Sasso underground laboratory}",
    eprint = "1905.05512",
    archivePrefix = "arXiv",
    primaryClass = "physics.ins-det",
    doi = "10.1140/epjc/s10052-019-7247-9",
    journal = "Eur. Phys. J. C",
    volume = "79",
    number = "9",
    pages = "747",
    year = "2019"
}

@inproceedings{Heise:2022iaf,
    author = "Heise, Jaret",
    title = "{The Sanford Underground Research Facility}",
    booktitle = "{Snowmass 2021}",
    eprint = "2203.08293",
    archivePrefix = "arXiv",
    primaryClass = "hep-ex",
    month = "3",
    year = "2022"
}

@article{MAGIS-100-2021etm,
    author = "Abe, Mahiro and others",
    collaboration = "MAGIS-100",
    title = "{Matter-wave Atomic Gradiometer Interferometric Sensor (MAGIS-100)}",
    eprint = "2104.02835",
    archivePrefix = "arXiv",
    primaryClass = "physics.atom-ph",
    reportNumber = "FERMILAB-PUB-21-031-AD-DI-FESS-QIS-T",
    doi = "10.1088/2058-9565/abf719",
    journal = "Quantum Sci. Technol.",
    volume = "6",
    number = "4",
    pages = "044003",
    year = "2021"
}

@misc{Heinzeldft,
author = {Heinzel, Gerhard and Rüdiger, Albrecht and Schilling, R},
year = {2002},
title = "{Spectrum and spectral density estimation by the Discrete Fourier transform (DFT), including a comprehensive list of window functions and some new flat-top windows}",
journal={Max Plank Institut f\"ur Gravitationsphysik},
howpublished={\href{https://hdl.handle.net/11858/00-001M-0000-0013-557A-5}{Max Plank Institut f\"ur Gravitationsphysik}}
}

@misc{bartington_instruments_magnetometer_2024,
	type = {Datasheet},
	title = {Magnetometer {Power} {Supplies}},
	howpublished = {\href{https://www.bartingtondownloads.com/wp-content/uploads/DS2520.pdf}{https://www.bartingtondownloads.com/wp-content/uploads/DS2520.pdf}},
	number = {DS2520/7},
	urldate = {2025-06-30},
	institution = {Bartington Instruments},
	author = {{Bartington Instruments}},
	month = may,
	year = {2024},
}

@misc{bartington_instruments_mag-13_2024,
	type = {Datasheet},
	title = {Mag-13 {Three}-{Axis} {Magnetic} {Field} {Sensors}},
	howpublished = {\href{https://www.bartingtondownloads.com/wp-content/uploads/DS3143.pdf}{https://www.bartingtondownloads.com/wp-content/uploads/DS3143.pdf}},
	number = {DS3143/21},
	urldate = {2025-06-30},
	institution = {Bartington Instruments},
	author = {{Bartington Instruments}},
	month = dec,
	year = {2024},
}

@techreport{measurement_computing_corporation_16-bit_202,
	type = {Datasheet},
	title = {16-bit {Voltage} {Measurement} {DAQ} {HAT} for {Raspberry} {Pi}},
	language = {en},
	number = {DS-MCC-128},
	institution = {Digilent Inc.},
	author = {{Measurement Computing Corporation}},
	month = dec,
	year = {2024},
}

@article{jossen_orion-triebzuge_2024,
	title = {Orion-{Triebzüge} für die {Matterhorn} {Gotthard}-{Bahn}},
	volume = {2024},
	url = {https://www.bahn-journalist.ch/pdf/ser_2024_05_mgb-orion.pdf},
	number = {5},
	journal = {\href{https://www.bahn-journalist.ch/pdf/ser_2024_05_mgb-orion.pdf}{Schweizer Eisenbahn-Revue}},
	author = {Jossen, Dario and Harbeke, Christian and Brauchli, René},
	year = {2024},
}

@misc{systemaufgaben_kundeninformation_fahrplanfeld_2024,
	title = {Fahrplanfeld 142: {Visp} - {Brig} - {Oberwald} - {Andermatt} - {Göschenen} (2025)},
	howpublished = {\href{https://widgets.oev-info.ch/publikation/jahresfpl/142.pdf}{https://widgets.oev-info.ch/publikation/jahresfpl/142.pdf}},
	urldate = {2025-11-21},
	publisher = {SBB Infrastruktur},
	author = {{Systemaufgaben Kundeninformation}},
	month = oct,
	year = {2024},
}

@misc{systemaufgaben_kundeninformation_fahrplanfeld_2024-1,
	title = {Fahrplanfeld 1982: {Autoverlad} {Furka} {Oberwald} - {Realp} (2025)},
	howpublished = {\href{https://www.oev-info.ch/sites/default/files/fap/2025/pdf/1982.pdf}{https://www.oev-info.ch/sites/default/files/fap/2025/pdf/1982.pdf}},
	urldate = {2025-11-21},
	publisher = {SBB Infrastruktur},
	author = {{Systemaufgaben Kundeninformation}},
	month = aug,
	year = {2024},
}

@misc{swiss_seismological_service_swiss_2016,
	title = {The {Swiss} {Seismological} {Service} ({SED})},
	howpublished = {\href{http://seismo.ethz.ch/export/sites/sedsite/knowledge/.galleries/pdf_brochures/Flyer_SED_EN.pdf_2063069299.pdf}{http://seismo.ethz.ch/export/sites/sedsite/knowledge\allowbreak/.galleries/pdf\_brochures/Flyer\_SED\_EN.pdf\_2063069299.pdf}},
	urldate = {2025-11-21},
	publisher = {ETH Zürich},
	author = {{Swiss Seismological Service}},
	year = {2016},
}

@article{peterson_observations_1993,
	title = {Observations and modeling of seismic background noise},
	url = {https://pubs.usgs.gov/publication/ofr93322},
	language = {en},
	number = {93-322},
	urldate = {2025-11-21},
	journal = {U.S. Geological Survey},
	author = {Peterson, Jon R.},
	year = {1993},
	doi = {10.3133/ofr93322},
	note = {ISSN: 2331-1258
Publication Title: Open-File Report},
}

@misc{magnetic_shield_corp_zg-3_2013,
	title = {{ZG}-3 {Zero} {Gauss} {Chambers}},
	howpublished = {\href{https://store-w4rbnih33r.mybigcommerce.com/content/ZG-3-brochure-web\%281\%29\%281\%29-REV-3-17-15.pdf}{https://store-w4rbnih33r.mybigcommerce.com/content/ZG-3-brochure-web\%281\%29\%281\%29-REV-3-17-15.pdf}},
	urldate = {2025-11-21},
	author = {{Magnetic Shield Corp.}},
	year = {2013},
}

@article{Araujo:2022kip,
    author = "Araujo, Gabriela R. and Baudis, Laura and Biondi, Yanina and Bismark, Alexander and Galloway, Michelle",
    title = "{The upgraded low-background germanium counting facility Gator for high-sensitivity {\ensuremath{\gamma}}-ray spectrometry}",
    eprint = "2204.12478",
    archivePrefix = "arXiv",
    primaryClass = "physics.ins-det",
    doi = "10.1088/1748-0221/17/08/P08010",
    journal = "JINST",
    volume = "17",
    number = "08",
    pages = "P08010",
    year = "2022"
}

@article{Baudis:2011am,
    author = "Baudis, L. and others",
    title = "{Gator: a low-background counting facility at the Gran Sasso Underground Laboratory}",
    eprint = "1103.2125",
    archivePrefix = "arXiv",
    primaryClass = "astro-ph.IM",
    doi = "10.1088/1748-0221/6/08/P08010",
    journal = "JINST",
    volume = "6",
    pages = "P08010",
    year = "2011"
}

@article{MCNPX-Polimi-1,
    author = {E.C. Miller and S.D. Clarke and M. Flaska and S.A. Pozzi and E. Padovani},
    title = "{MCNPX-PoliMi Post-Processing Algorithm for Detector Response Simulation}",
    journal = {\href{https://resources.inmm.org/system/files/jnmm/vol_40/V-40_2.pdf}{Journal of Nuclear Materials Management}},
    year = {2012},
    volume = {40},
    issue = {2},
    pages = {34 - 41},
    url = {https://resources.inmm.org/system/files/jnmm/vol_40/V-40_2.pdf}
}

@article{CLARKE2013135,
title = "{Verification and validation of the MCNPX-PoliMi code for simulations of neutron multiplicity counting systems}",
journal = {Nucl. Instrum. Methods Phys. Res. A},
volume = {700},
pages = {135-139},
year = {2013},
issn = {0168-9002},
doi = {https://doi.org/10.1016/j.nima.2012.10.025},
url = {https://www.sciencedirect.com/science/article/pii/S0168900212011576},
author = {S.D. Clarke and E.C. Miller and M. Flaska and S.A. Pozzi and R.B. Oberer and L.G. Chiang}
}

@article{MCNPX,
    author = {L.S. Waters and others},
    title = "{The MCNPX Monte Carlo Radiation Transport Code}",
    journal = {AIP Conference Proceedings},
    volume = {896},
    year = {2007},
    pages = {81-90},
    doi = {10.1063/1.2720459}
}

@article{RLemrani_2006,
    author = "Lemrani, R. and Gerbier, G.",
    collaboration = "EDELWEISS",
    title = "{Update of neutron studies in EDELWEISS}",
    doi = "10.1088/1742-6596/39/1/033",
    journal = "J. Phys. Conf. Ser.",
    volume = "39",
    pages = "145--147",
    year = "2006"
}

@article{CHOI2019263,
title = "{Study of $n/\gamma$ discrimination using 3He proportional chamber in high gamma-ray fields}",
journal = {Nuclear Engineering and Technology},
volume = {51},
number = {1},
pages = {263-268},
year = {2019},
issn = {1738-5733},
doi = {https://doi.org/10.1016/j.net.2018.08.013},
url = {https://www.sciencedirect.com/science/article/pii/S1738573318301050},
author = {J. Choi and J. Park and J. Son and Y.K. Kim}
}

@article{rast2022geology,
  title="{Geology along the Bedretto tunnel: kinematic and geochronological constraints on the evolution of the Gotthard Massif (Central Alps)}",
  author={Rast, Markus and Galli, Andrea and Ruh, Jonas B and Guillong, Marcel and Madonna, Claudio},
  journal={Swiss Journal of Geosciences},
  volume={115},
  number={1},
  pages={8},
  year={2022},
  publisher={Springer},
  doi={10.1186/s00015-022-00409-w}
}

@misc{LND,
  author = {\relax LND Inc.},
  title = "{252108: Cylindrical He3 neutron detector}",
  howpublished = {\href{https://www.lndinc.com/products/neutron-detectors/252108/}{https://www.lndinc.com/products/neutron-detectors/252108/}},
  note = {https://www.lndinc.com/products/neutron-detectors/252108/}
}

@misc{Ortec,
  author = {\relax Ortec Amptek},
  title = "{142PC Preamplifier}",
  howpublished = {\href{https://www.ortec-online.com/products/electronic-instruments/preamplifiers/142pc}{https://www.ortec-online.com/products/electronic-instruments/preamplifiers/142pc}},
  note = {https://www.ortec-online.com/products/electronic-instruments/preamplifiers/142pc}
}

@misc{CAEN-DPP-PHA,
    author={CAEN},
    title="{Digital Pulse Processing for the Pulse Height Analysis}",
    howpublished={\href{https://www.caen.it/products/dpp-pha/}{https://www.caen.it/products/dpp-pha/}},
    note = {https://www.caen.it/products/dpp-pha/}
}

@misc{CAEN5724,
  author = {CAEN},
  title = "{DT5724 2/4 Channel 14 bit 100 MS/s Digitizer}",
  howpublished = {\href{https://www.caen.it/products/dt5724/}{https://www.caen.it/products/dt5724/}},
  note = {https://www.caen.it/products/dt5724/}
}

@misc{CAEN5730,
  author = {CAEN},
  title = "{DT5730S 8 Channel 14 bit 500 MS/s Digitizer}",
  howpublished = {\href{https://www.caen.it/products/dt5730/}{https://www.caen.it/products/dt5730/}},
  note = {https://www.caen.it/products/dt5730/}
}

@misc{CAEN5533E,
  author = {CAEN},
  title = "{DT5533E 4 Channel 4 kV/3 mA (4 W) Desktop HV Power Supply Module}",
  howpublished = {\href{https://www.caen.it/products/dt5533e/}{https://www.caen.it/products/dt5533e/}},
  note = {https://www.caen.it/products/dt5533e/}
}

@misc{CAENsipm,
    author = {CAEN},
    title = "{Single Channel 85 V/10 mA Power Supply Module for SiPM}",
    url = {https://www.caen.it/subfamilies/up-to-85-v-power-supply-module-for-sipm-a7585d/},
    howpublished = {\href{https://www.caen.it/subfamilies/up-to-85-v-power-supply-module-for-sipm-a7585d/}{https://www.caen.it/subfamilies/up-to-85-v-power-supply-module-for-sipm-a7585d/}}
}

@misc{Bedretto,
title = {Bedretto Underground Laboratory for Geosciences and Geoenergies},
author = {{\relax ETH Z\"urich}},
howpublished={\href{https://www.bedrettolab.ethz.ch/home/}{https://www.bedrettolab.ethz.ch/home/}}
}

@misc{EJ-200,
title = "{EJ-200, EJ-204, EJ-208, EJ-212 - Plastic Scintillators}",
howpublished={\href{https://eljentechnology.com/products/plastic-scintillators/ej-200-ej-204-ej-208-ej-212}{https://eljentechnology.com/products/plastic-scintillators/ej-200-ej-204-ej-208-ej-212}},
author = {\relax Eljen Technology}
}

@article{arduiniLongBaselineAtomInterferometer2023,
    author = "Arduini, G. and others",
    title = "{A Long-Baseline Atom Interferometer at CERN: Conceptual Feasibility Study}",
    eprint = "2304.00614",
    archivePrefix = "arXiv",
    primaryClass = "physics.atom-ph",
    reportNumber = "CERN-PBC-REPORT-2023-002, CERN-TH-2023-051",
    month = "4",
    year = "2023"
}

@article{mitchellMAGIS100EnvironmentalCharacterization2022,
  title = {{{MAGIS-100}} Environmental Characterization and Noise Analysis},
  author = {Mitchell, J. and Kovachy, T. and Hahn, S. and Adamson, P. and Chattopadhyay, S.},
  year = {2022},
  month = jan,
  journal = {JINST},
  volume = {17},
  number = {01},
  pages = {P01007},
  publisher = {IOP Publishing},
  issn = {1748-0221},
  doi = {10.1088/1748-0221/17/01/P01007},
  urldate = {2025-05-30},
  langid = {english}
}

@article{korun_martincic_pucelj_ravnik_1996, 
    title={Measurement of the ambient neutron background with a high-resolution $\gamma$-ray spectrometer}, 
    author={M. Korun and R. Martincic and B. Pucelj and M. Ravnik}, year={1996}, 
    month={Jan}, 
    pages={p. 116–118},
    journal={Symposium on radiation protection in neighbouring countries in Central Europe - 1995. Proceedings},
    eprint={https://inis.iaea.org/records/wmwz9-m0748}
}

@article{Gorshkov1964,
    author = {G.V. Gorshkov and V.A. Zyabkin and O.S. Tsvetkov},
    title = {The neutron background at the surface of the earth},
    journal = {Soviet Atomic Energy},
    year = {1964},
    volume = {17},
    issue = {6},
    pages = {1256 - 1260},
    doi={10.1007/BF01122773}
}

@misc{swisstopo, 
title={swissALTI3D}, 
howpublished={\href{https://www.swisstopo.admin.ch/en/height-model-swissalti3d}{https://www.swisstopo.admin.ch/en/height-model-swissalti3d}}, 
author={\relax Federal Office of Topography swisstopo} 
}

@article{nasa2019aster,
  title="{ASTER global digital elevation model V003}",
  author={\relax{NASA/METI/AIST/Japan Spacesystems and US/Japan ASTER Science Team, K}},
  journal={NASA EOSDIS Land Process. DAAC},
  year={2019},
  note={Date Accessed: 2025-06-24},
  doi={10.5067/ASTER/ASTGTM.003}
}

@article{Haffke:2011fp,
    author = "Haffke, M. and others",
    title = "{Background Measurements in the Gran Sasso Underground Laboratory}",
    eprint = "1101.5298",
    archivePrefix = "arXiv",
    primaryClass = "astro-ph.IM",
    doi = "10.1016/j.nima.2011.04.027",
    journal = "Nucl. Instrum. Meth. A",
    volume = "643",
    pages = "36--41",
    year = "2011"
}

@article{Ascenzo:2025ujf,
    author = "Ascenzo, Lorenzo and Benato, Giovanni and Chu, Yingjie and Di Carlo, Giuseppe and Molinario, Andrea and Vernetto, Silvia",
    title = "{Characterization of a GAGG detector for neutron measurements in underground laboratories}",
    eprint = "2504.16889",
    archivePrefix = "arXiv",
    primaryClass = "physics.ins-det",
    doi = "10.1140/epjc/s10052-025-14807-5",
    journal = "Eur. Phys. J. C",
    volume = "85",
    number = "9",
    pages = "1057",
    year = "2025"
}

@article{Wulandari:2003cr,
    author = "Wulandari, H. and Jochum, J. and Rau, W. and von Feilitzsch, F.",
    title = "{Neutron flux underground revisited}",
    eprint = "hep-ex/0312050",
    archivePrefix = "arXiv",
    doi = "10.1016/j.astropartphys.2004.07.005",
    journal = "Astropart. Phys.",
    volume = "22",
    pages = "313--322",
    year = "2004"
}

@article{HASHEMINEZHAD1998100,
title = "{Limitation on the response of 3He counters due to intrinsic alpha emission}",
journal = {Nucl. Instrum. Methods Phys. Res. A.},
volume = {416},
number = {1},
pages = {100-108},
year = {1998},
issn = {0168-9002},
doi = {https://doi.org/10.1016/S0168-9002(98)00565-8},
url = {https://www.sciencedirect.com/science/article/pii/S0168900298005658},
author = {S.R Hashemi-Nezhad and L.S Peak}
}

@article{debickiLNGS, 
    author = "Debicki, Z. and others",
    editor = {Capdevielle, Jean-No{\"e}l and Engel, Ralph and Pattison, Bryan},
    title = "{Thermal neutrons at Gran Sasso}",
    doi = "10.1016/j.nuclphysbps.2009.09.084",
    journal = "Nucl. Phys. B Proc. Suppl.",
    volume = "196",
    pages = "429--432",
    year = "2009"
}

@article{CaptureGatedNeutron,
    author = {T. Shi and J. Nattress and M. Mayer and M.W. Lin and I. Jovanovic},
    year = {2016},
    month = {09},
    pages = {},
    title = "{Neutron spectroscopy by thermalization light yield measurement in a composite heterogeneous scintillator}",
    volume = {839},
    journal = {Nucl. Instrum. Methods Phys. Res. A},
    doi = {10.1016/j.nima.2016.09.041}
}

@article{ConcreteShielding,
    author = {S. Barbhuiya and B.B. Das and P. Norman and T. Qureshi },
    title = "{A comprehensive review of radiation shielding concrete: Properties, design, evaluation, and applications}",
    journal = {Structural Concrete},
    year = {2024},
    volume={26},
    issue={2}, 
    doi={10.1002/suco.202400519},
    url={https://doi.org/10.1002/suco.202400519}
}

@misc{rad7,
    author={Durridge},
    title="{RAD7 radon detector}",
    howpublished = {\href{https://durridge.com/products/rad7-radon-detector/}{https://durridge.com/products/rad7-radon-detector/}},
    url = {https://durridge.com/products/rad7-radon-detector/}
}

@misc{drystik,
    author={Durridge},
    title="{Drystik Model ADS-3}",
    howpublished = {\href{https://durridge.com/products/drystik/}{https://durridge.com/products/drystik/}},
    url = {https://durridge.com/products/drystik/}
}

@article{Scovell:2023pxo,
    author = "Scovell, P. R. and Meehan, E. and Paling, S. M. and Thiesse, M. and Liu, X. and Ghag, C. and Ginsz, M. and Quirin, P. and Ralet, D.",
    title = "{Ultra-low background germanium assay at the Boulby Underground Laboratory}",
    eprint = "2308.03444",
    archivePrefix = "arXiv",
    primaryClass = "physics.ins-det",
    doi = "10.1088/1748-0221/19/01/P01017",
    journal = "JINST",
    volume = "19",
    number = "01",
    pages = "P01017",
    year = "2024"
}

@article{ampollini2023sub,
  title="{Sub-background radiation exposure at the LNGS underground laboratory: dosimetric characterization of the external and underground facilities}",
  author={Ampollini, Marco and others},
  journal={Frontiers in Physics},
  volume={11},
  pages={1274110},
  year={2023},
  publisher={Frontiers Media SA},
  doi={10.3389/fphy.2023.1274110}
}

@article{LSMgamma,
    author = {D. Malczewski and J. Kisiel and J. Dorda},
    title = {Gamma background measurements in the Laboratoire Souterrain de Modane},
    journal = {J. Radioanal. Nucl. Chem.},
    year = {2012},
    volume={292},
    issue={2},
    pages={751-756},
    url={https://doi.org/10.1007/s10967-011-1497-9},
    doi={10.1007/s10967-011-1497-9}
}

@article{LNGSfast,
    author = "Bertoni, R. and others",
    title = "{Measurement of fast neutron background in the Gran Sasso underground laboratory using a geyser-concept bubble-chamber}",
    doi = "10.1140/epjc/s10052-023-11522-x",
    journal = "Eur. Phys. J. C",
    volume = "83",
    number = "5",
    pages = "354",
    year = "2023"
}

@article{GEANT4:2002zbu,
    author = "Agostinelli, S. and others",
    collaboration = "GEANT4",
    title = "{GEANT4 - A Simulation Toolkit}",
    reportNumber = "SLAC-PUB-9350, FERMILAB-PUB-03-339, CERN-IT-2002-003",
    doi = "10.1016/S0168-9002(03)01368-8",
    journal = "Nucl. Instrum. Meth. A",
    volume = "506",
    pages = "250--303",
    year = "2003"
}

@article{Cirrone:2010zz,
    author = "Cirrone, G. A. P. and Cuttone, G. and Di Rosa, F. and Pandola, L. and Romano, F. and Zhang, Q.",
    title = "{Validation of the Geant4 electromagnetic photon cross-sections for elements and compounds}",
    doi = "10.1016/j.nima.2010.02.112",
    journal = "Nucl. Instrum. Meth. A",
    volume = "618",
    pages = "315--322",
    year = "2010"
}

@article{bedretto_charaterization,
AUTHOR = {Ma, X. and others},
TITLE = "{Multi-disciplinary characterizations of the BedrettoLab -- a new underground geoscience research facility}",
JOURNAL = {Solid Earth},
VOLUME = {13},
YEAR = {2022},
NUMBER = {2},
PAGES = {301--322},
URL = {https://se.copernicus.org/articles/13/301/2022/},
DOI = {10.5194/se-13-301-2022}
}

@phdthesis{Tarka_2012, 
    title={Studies of Neutron Flux Suppression from a $\gamma$-ray Source and The GERDA Calibration System}, 
    school={\href{https://doi.org/10.5167/uzh-74790}{University of Zurich}},
    author={Tarka, Michal}, 
    year={2012}, 
    month={May}
}

\pagebreak
\begin{appendices}
    
\section{Gator Measurement}\label{secA1}

\begin{figure}[h]
    \centering
    \includegraphics[width=0.45\textwidth]{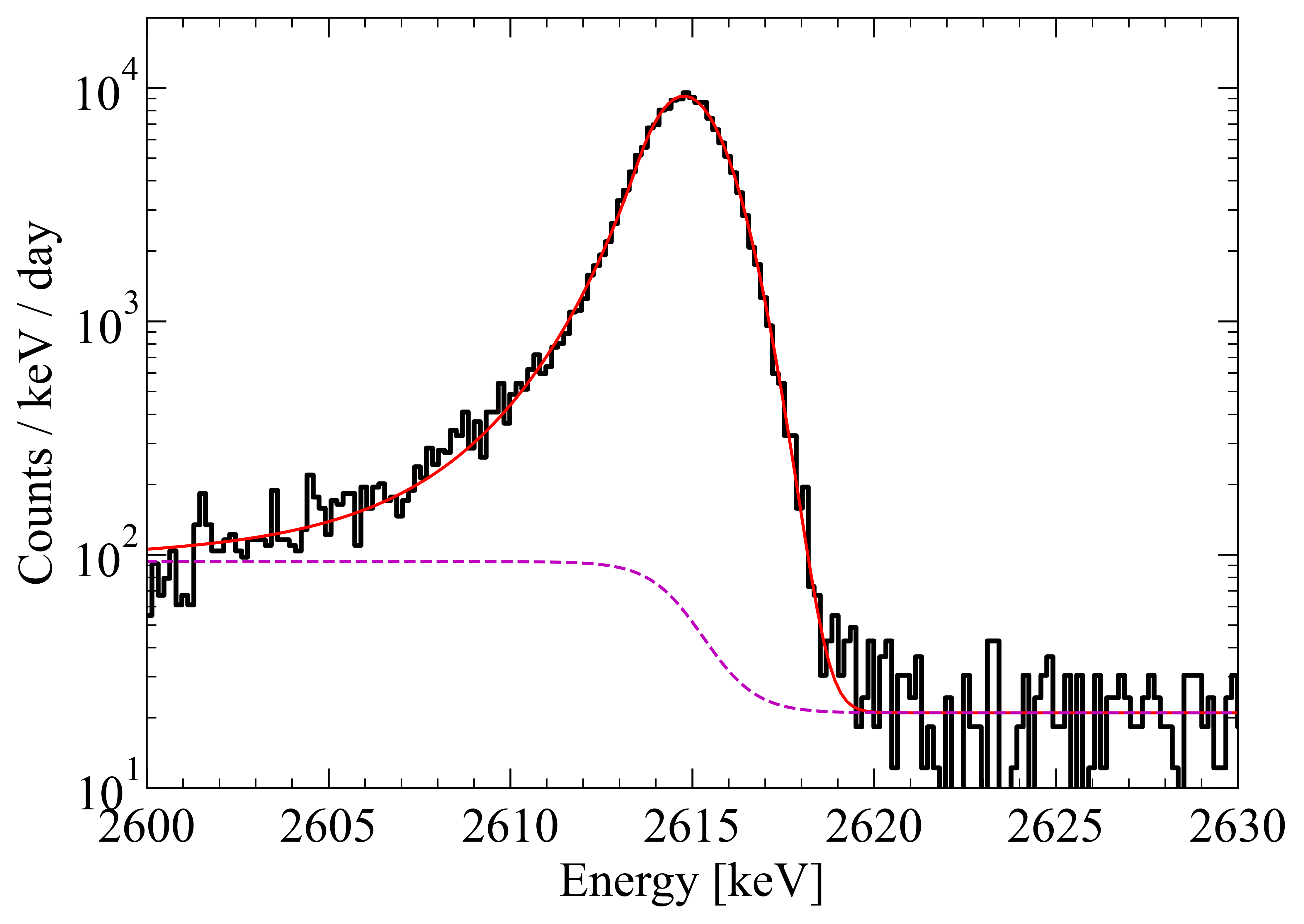}
    \caption{2614.5 keV $^{208}$Tl peak fitted using a Crystall ball peak plus a sigmoid step background. Fit shown in red with background step component shown in magenta. Data (black) corresponds to the TM2500 granite sample.}\label{fig:2615_peak_fit}
    \end{figure}

    To determine the activity, the observed gamma-ray lines in the spectra are analyzed by fitting the gamma peaks using a Crystallball peak fit with a step sigmoid background. Then, a peak window of $\pm 3-5 \sigma$ is chosen depending on the extent of the low-energy tail of the peak. A background window of the same size is also chosen on both the low- and high-energy side of the peak. The difference gives the net counts, which are used for further computations. An example is shown in Fig.~\ref{fig:2615_peak_fit}.
    
    Using the net counts of the peak obtained after background subtraction, the specific activity of each gamma line can be computed using Eq.~\ref{eq:act_rock} where $\epsilon$ is the efficiency of the gamma line obtained from the GEANT4 simulation of the sample in Gator, t is the measurement time (in seconds), BR is the branching ratio of the gamma line, and m is the mass of the sample, 1.4 kg in the case of the one from TM2500.
    
    \begin{equation}
            \text{A (Bq/kg)} = \frac{\text{Net Counts}}{\epsilon \cdot \text{t} \cdot \text{BR} \cdot \text{m}}
            \label{eq:act_rock}
    \end{equation}

    Systematic uncertainties include a 3~\% geometric uncertainty for the relative position of the rock sample to the HPGe crystal as well as a 4~\% uncertainty for cross sections within the Geant4 libraries \cite{GEANT4:2002zbu,Cirrone:2010zz}. Statistical uncertainty is estimated utilizing a Monte-Carlo sampling of the peak and include Poisson errors. Uncertainties from the branching ratio were included but contribute less than 0.1~\% of the total uncertainty.

    Table~\ref{tab:rock_activity} summarises the activity and uncertainty for each measured isotope for the TM2500 rock sample within the $^{235}$U, $^{238}$U, $^{232}$Th, and $^{40}$K decay chains. The gamma lines included within the analysis were chosen for prominence within the spectrum and those with high branching ratios. Individual isotopic activities are averaged with an error weighting to compute the values shown in Table~\ref{tab:rock_activity_final}.

\begin{table}
        \centering
        \begin{tabular}{cccc}
        \hline\noalign{\smallskip}
        \textbf{Energy} & \multirow{2}{*}{\textbf{Parent}} & \textbf{Decay} & \textbf{Activity} \\
        \textbf{(keV)} &  & \textbf{Chain} & \textbf{(Bq/kg)} \\
        \hline
        185.7  & $^{235}$U  & $^{235}$U  & 10.0 $\pm$ 0.6  \\
        186.2  & $^{226}$Ra & $^{238}$U  &  161.3 $\pm$ 3.8 \\
        238.6  & $^{212}$Pb & $^{232}$Th & 136.9 $\pm$ 6.3  \\
        295.2  & $^{214}$Pb & $^{238}$U  & 172.0 $\pm$ 8.9  \\
        338.6  & $^{228}$Ac & $^{232}$Th & 107.8 $\pm$ 5.9  \\
        351.9  & $^{214}$Pb & $^{238}$U  & 173.4 $\pm$ 9.2  \\
        583.2  & $^{208}$Tl & $^{232}$Th & 120.8 $\pm$ 5.4  \\
        609.3  & $^{214}$Bi & $^{238}$U  & 157.5 $\pm$ 7.9  \\
        727.3  & $^{212}$Bi & $^{232}$Th & 136.2 $\pm$ 10.7  \\
        860.5  & $^{208}$Tl & $^{232}$Th & 117.9 $\pm$ 6.2  \\
        911.2  & $^{228}$Ac & $^{232}$Th & 119.4 $\pm$ 6.2  \\
        1001.0 & $^{234}$Pa & $^{238}$U  & 220.6 $\pm$ 18.0  \\
        1120.3 & $^{214}$Bi & $^{238}$U  & 156.1 $\pm$ 8.2  \\
        1460.8 & $^{40}$K   & $^{40}$K   & 1125.1 $\pm$ 73.0  \\
        1764.5 & $^{214}$Bi & $^{238}$U  & 150.3 $\pm$ 9.1  \\
        2614.5 & $^{208}$Tl & $^{232}$Th & 112.5 $\pm$ 5.2  \\
        \noalign{\smallskip}\hline
        \end{tabular}
        \caption{The specific activity obtained from each of the major gamma-ray peaks for the granite sample from TM2500 are listed in this table.}
        \label{tab:rock_activity}
\end{table}

\section{Time variation of the radon concentration}\label{app:radon_evolution}

The RAD7 radon monitor electrostatically collects positively charged radon progeny. Water vapour in the air neutralises the ions, which diminishes the collection efficiency. Therefore, the relative humidity inside the detector needs to be monitored and reduced. A laboratory gas drying unit with a desiccant is typically used. To cope with the high humidity in the Bedretto tunnel, a Nafion\textsuperscript{TM}-based humidity exchanger (Durridge Drystick) is added, to reduce the desiccant consumption rate during long-term measurements.

\begin{figure*}[h]
    \centering
    \includegraphics[width=\linewidth]{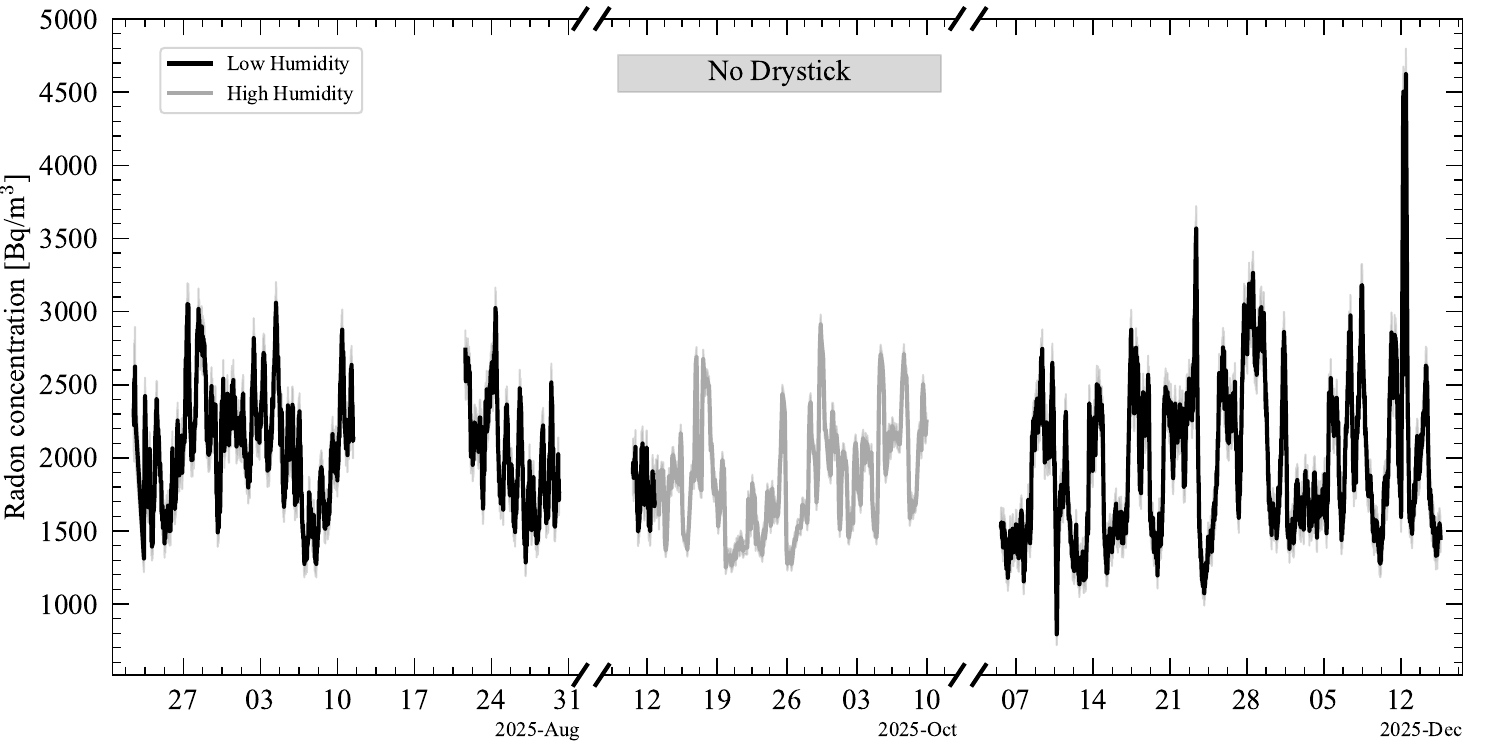}
    \caption{Time evolution of the radon concentration in the unventilated section at TM3500 as measured using a Durridge RAD7 electrostatic radon monitor.}
    \label{fig:radon_time_evolution}
\end{figure*}

Figure~\ref{fig:radon_time_evolution} shows the evolution of the measured radon concentration during the period between end of July until middle of December 2025. At the end of August 2025, the Drystick’s internal pump malfunctioned, rendering subsequent data unreliable. Before its replacement, the monitor was operated only with the laboratory drying unit, initially maintaining the air at 4\% relative humidity. After a few days, signs of saturation became evident, resulting in a rapid humidity increase. Enlarging the measurement cycle to 2 hours allowed to stabilise the humidity at about 60\% until mid-October, after which further a increase in humidity required exclusion of subsequent data.
Following the installation of a new Drystick module in early November 2025, stable data taking with a relative humidity below 7\% could be continued.

\section{Overburden distribution}
\begin{figure}[h]
    \centering
    \includegraphics[width=\linewidth]{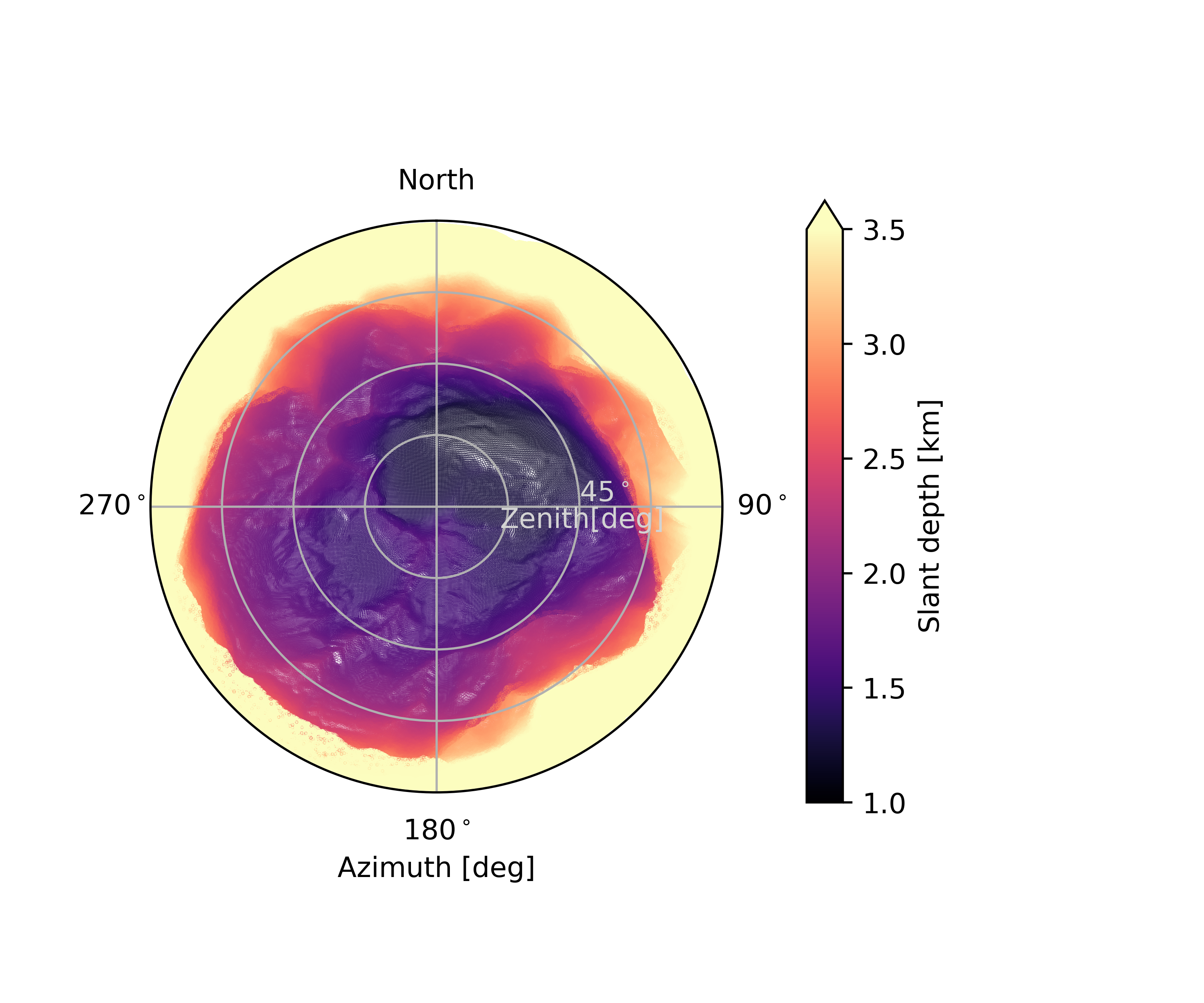}
    \caption{Slant depth to TM3500 as function of azimuth and zenith angle. The shallowest overburden is $1.16~\mathrm{km}$ to the North-East.}
    \label{fig:overburdengeometry}
\end{figure}

The geometric overburden above TM3500 was computed using topological data from the Bundesamt für Landestopografie swisstopo. Figure~\ref{fig:overburdengeometry} shows the slant depth from the surface to TM3500 as function of angles. 

\end{appendices}

\end{document}